\documentclass[smallextended]{svjour3}       % onecolumn (second format)
\pdfoutput=1
\usepackage{graphicx}
\usepackage{caption}
\usepackage{subcaption}
\usepackage{hyperref}
\usepackage[linesnumbered,ruled,vlined,commentsnumbered]{algorithm2e}
\usepackage{latexsym}
\usepackage{booktabs} % For formal tables
\usepackage{adjustbox}
\usepackage{color, colortbl,xcolor}

\begin{document}

\title{Using Core-Periphery Structure to Predict\\ High Centrality Nodes in Time-Varying Networks}

\titlerunning{Predicting Centrality in Temporal Networks}        % if too long for running head

\author{Soumya Sarkar \and
        Sandipan Sikdar \and Sanjukta Bhowmick \and Animesh Mukherjee  %etc.
}

\authorrunning{Sarkar et al} % if too long for running head

\institute{Soumya Sarkar \at
              \email{soumya015@iitkg.ac.in}           
           \and
              Sandipan Sikdar \at
             \email{sandipansikdar@cse.iitkgp.ernet.in}
           \and
             Sanjukta Bhowmick \at
             \email{sbhowmick@unomaha.edu}
             \and
             Animesh Mukherjee \at
             \email{animeshm@cse.iitkgp.ernet.in}        
}

\date{Received: date / Accepted: date}

%\begin{document}
\maketitle

\begin{abstract}
Vertices with high betweenness and closeness centrality represent influential entities in a network. An important problem for time varying networks is to know a-priori, using minimal computation, whether the influential vertices of the current time step will retain their high centrality, in the future time steps, as the network evolves. 

In this paper, based on empirical evidences from several large real world time varying networks, we discover a certain class of networks \if{0}which we term as {\em  Core Connected} (CC)\fi where the highly central vertices are part of the innermost core of the network and this property is maintained over time. As a key contribution of this work, we propose \textit{novel heuristics} to identify these networks in an optimal fashion and also develop {\em a two-step algorithm for predicting high centrality vertices}.  \iffalse{}Specifically, the point we wish to stress here is that if (i) the highly central vertices are part of the innermost core in a network and (ii) this set of vertices in the innermost core do not change significantly, then the highly central vertices remain unchanged over time.\fi Consequently, we show for the first time that for such networks, expensive shortest path computations in each time step as the network changes can be completely avoided; instead we can use time series models (e.g., ARIMA as used here) to predict the overlap between the high centrality vertices in the current time step to the ones in the future time steps. Moreover, once the new network is available in time, we can find the high centrality vertices in the top core simply based on their high degree. 

To measure the effectiveness of our framework, we perform prediction task on a large set of diverse time-varying networks. We obtain $F1$-scores as high as \textbf{0.81} and \textbf{0.72} 
in predicting the top $m$ closeness and betweenness centrality vertices respectively for real networks where the highly central vertices mostly reside in the innermost core. For synthetic networks that conform to this property we achieve $F1$-scores of \textbf{0.94} and \textbf{0.92} for closeness and betweenness respectively. We validate our results by showing that the practical effects of our predicted vertices match the effects of the actual high centrality vertices. Finally, we also provide a formal sketch demonstrating why our method works.

%To the best of our knowledge this is the {\em first algorithm to  predict the ids of the highly central vertices} rather than simply predicting the average centrality of the network. \textit{Our method allows users to distinguish between networks where prediction is possible from those where it is not}. Together, these novel features promote better understanding  and more accurate prediction of high centrality nodes in time-varying networks.


\end{abstract}

\section{Introduction}
One of the important problems in time-varying networks is predicting how their features change with time. If this information is known a-priori using minimal computation, then users can take appropriate action in advance to utilize such features. The most significant among network properties are the centrality features, that are used to estimate the importance of a vertex in a network.  

Information can spread more quickly when high closeness centrality vertices are selected as the initial seeds. Similarly, vaccinating high betweenness centrality vertices, through which most of the shortest paths pass, can reduce the spread of disease. 
The central vertices also play an important role in spreading influence in a social network as has been observed in several  works~\cite{carnes2007maximizing,morselli2013predicting,liu2018identifying,agneessens2017geodesic}. In a dynamic setting (where the network changes over time) knowing these highly central vertices beforehand is of prime importance as it would facilitate in developing strategies for targeted advertising or setting up infrastructure for vaccination drives. However, this might result in expensive re-computation of shortest paths as the network varies over time. Our goal is to  develop algorithms based on the network structure,  so that such re-computations are avoided.

Current approaches focus on predicting the average centrality values of the network~\cite{kim2012centrality}. However, note that
most applications, such as the ones discussed above, require the ids of {\em only the top-$k$ centrality vertices}, not the values or the ranking of {\em all} the vertices in the network. Therefore, simply predicting the average centrality over the entire network may not be useful in a majority of the practical contexts.
 
In this paper, we {\bf present a two-step algorithm for predicting the high centrality vertices of time-varying networks}. In the first step, we predict the overlap between the set of high centrality vertices\footnote{In this paper, when we mention highly central  vertices, we specifically refer to high closeness or betweenness centrality and not other types of centralities.} of the current time step to the set of high centrality vertices of the future time step. In the next step, assuming that the network snapshot is already available in time, we analyze its innermost core to find the ids of the high centrality vertices.

\noindent{\bf The rationale for our prediction method}: The key to our prediction method is based on a unique \textit{hypothesis} that in many real world time-varying networks, \textit{most of the highly central vertices reside in the innermost core}. In other words, a \textit{large fraction of the shortest paths connecting pairs of vertices in such networks pass through the innermost core}; the vertices in the periphery (and the outer shells) of the network are mostly connected via the vertices residing in the innermost core of the network. A key contribution of our work is that we develop a set of \textit{novel heuristics to classify networks based on the extent to which the highly central vertices are in the innermost core}. Our heuristics do not require the explicit computation of the centrality values. We also separately report our predictions for each class of networks. We empirically demonstrate that the higher the number of high centrality vertices in the inner core, the higher is the accuracy with which we can predict these  vertices for future time steps. For real networks that maintain this property to the largest extent, our $F1$-score for prediction is \textbf{0.81} for closeness and \textbf{0.72} for betweenness. similarly, for synthetic networks that main this property to best extent, the $F1$-score for prediction is \textbf{0.94} for closeness and \textbf{0.92} for betweenness.

\if{0}on a special core-periphery property that we introduce, called the {\em core connectedness} ($CC$), which measures how closely the vertices in the outer cores are connected to the innermost core of the network (details in Section~\ref{theory}). We demonstrate that the more a network conforms to the CC property across the time steps, the higher the accuracy of our predictions.

We define the CC property to ensure that most of the paths representing shortest distance between vertices pass through the innermost core  
(refer to section ~\ref{theory}). This increases the likelihood that the high betweeness and closeness centrality vertices are also in the innermost core. This property facilitates the prediction of the vertices with these high centrality metrics. We provide empirical results of our algorithm on several real-world time-varying networks.\fi

\noindent{\bf Validation}: In addition to $F1$-scores, we further validate our results by comparing how the predicted and actual high centrality vertices perform in a practical context. For high closeness centrality vertices, we compare the time to spread a message when the high centrality vertices are taken as seeds, and for the high betweenness centrality vertices, we compare how the length of the diameter increases as the high betweenness centrality vertices are deleted from the network. For these experiments we select a set of random vertices as control, and compare the performance of the actual high centrality vertices, predicted high centrality vertices and the randomly selected vertices.  For networks where we could predict the results with high accuracy, the effect of the original and predicted vertices are very similar, and  these results are markedly different from the effect of the randomly selected vertices. Interestingly, for networks, where our prediction accuracy is low, the effect is similar for closeness/betweenness centrality for all the three sets of vertices. This result indicates that networks with low prediction accuracy do not have significantly high closeness/betweenness centrality vertices and therefore the prediction itself does not serve any practical purpose. 
To summarize, our {\bf key contributions} are;

 \vspace{-1mm}
 \begin{itemize}  
\item %Present a hypothesis that high centrality vertices of many real world time-varying networks, are in the innermost core of the network. 
Develop a set of heuristics to classify networks based on the extent to which the highly central vertices are part of the innermost core (section~\ref{sec:classification}). To perform the classification, these heuristics do not require the explicit computation of the high centrality vertices. 
\if{0}core connectedness (CC) property that allows us to efficiently predict the highly central vertices in time-varying networks (section \ref{theory}). Moreover, we  demonstrate the utility of core periphery structure of the network in obtaining the set of shortest distance based high centrality vertices in the network. 
  
\item Develop a classification method, based on the CC properties and the  overlap among the vertices in the innermost core across consecutive time steps, to distinguish between networks for which the high centrality vertices can be predicted with high accuracy and networks for which reliable prediction is not possible. ().\fi

\item Develop an efficient algorithm to predict the high betweenness and closeness centrality vertices. To the best of our knowledge, this is { the first algorithm to predict the vertices that does not require explicit computation of the centralities} (section~\ref{algorithm}). %and not just the average centrality value of the vertices in the network (section~\ref{algorithm}).

\item Validate our results in practical application scenarios, namely message spreading and increasing the diameter, to show that the effect of our predicted vertices is similar to the actual ones. For networks, where our prediction accuracy is low, we show that even a random selection of vertices can produce same results as the actual high closeness/ betweenness centrality vertices. This indicates that for such networks, there are no significantly high closeness/betweenness centrality vertices. Thus, in these cases, prediction of high closeness/betweenness centrality vertices is of no practical use.
\item Present a theoretical rationale for our  algorithm.

 \end{itemize}
 
Our condition for accurate prediction of high centrality vertices is that they are part of the innermost core of the network. 
As we shall see, in most real-world time-varying networks in our dataset, this condition holds. This condition is also true for a large number of synthetically generated networks. Even for the networks where the condition only holds to a moderate extent, our predictions are reasonably well. We also observe that for those networks where the condition fails to hold, it is not really worth finding high centrality vertices (by any algorithm) as there is no explicit ranking present in such networks. The functional effect of the high centrality vertices is equivalent to those of a randomly selected set of vertices.
 
\section{Related work}
We now present some of the research related to the topics discussed in this paper.

\noindent{\bf Core-periphery analysis}: Coreness of nodes and $K$-core subgraph network, introduced by Seidman~\cite{seidman1983network}, have been shown to have many applications. It has been used as a submodule in designing solutions to community detection problems \cite{peng2014accelerating} and in developing dissemination strategies \cite{kitsak2010identification}.  
Batagelj and Zaversnik \cite{batagelj2003m} proposed an algorithm that requires $O(max(|E|, |V|))$ runtime and $O(|E|)$ space to perform $K$-core decomposition. Several algorithms have been proposed focusing on the problem that web scale graphs have non trivial edge sizes~\cite{obrien2014locally,khaouid2015k,govindan2016nimblecore} and it is not always possible to load the entire list of edges in the memory of a single machine. In~\cite{barucca2016centrality} the authors generate synthetic networks using degree corrected stochastic blockmodel and investigate the relation between core-periphery structure and centrality. Although prior works have established various applications of core-periphery decomposition and its relation to centrality, we are the first to investigate the utility of node coreness in the context of temporal networks. We also propose a comprehensive strategy demonstrating how to utilize this connection between coreness and centrality in dynamic networks to compute the high centrality nodes in an inexpensive fashion.

\noindent{\bf Centrality in time-varying networks}: \textcolor{black}{ Time varying networks have been a topic of huge interest largely because in contrast to static networks, they take into account the time frame of pairwise interactions, which explains the dynamics of roles undertaken by nodes in the network.  We refer to comprehensive reviews~\cite{holme2012temporal,holme2015modern} for detailed coverage on the topic. One of the key directions of research in temporal networks has been identifying influential agents in a dynamic setting. In this lines, there have been two types of research -- the first type have attempted to adapt definitions of traditional centrality metrics to the dynamic setting and the second type have attempted to predict centrality of nodes from previous versions of the network, without explicitly calculating the metrics for the current version. }

\noindent\textcolor{black}{\textbf{Robustness of centrality measures}: One of the first studies which address issues related to
robustness of centrality measures corresponding to the minor variations in the network structure or noise has been reported by Borgatti et. al.~\cite{borgatti2006robustness}. The inherent dependence of centrality metrics on the network structure has been further studied by Braha et.al.~\cite{braha2006centrality,braha2009time,hill2010dynamic}. Their seminal work throws insights on the dynamical behavior, including a dramatic time dependence of the role of nodes apparent in the different snapshots of the network. These characteristics are not apparent in static (time aggregated) analysis of node connectivity and network topology thus motivating the necessity to have separate definitions of centrality in temporal networks. In~\cite{kim2012temporal,nicosia2013graph}, the authors utilize a powerful framework of time ordered graphs which transforms dynamic networks to static networks with directed flows. The authors adapt definitions of centrality such as closeness, betweenness and degree for temporal networks, by defining temporal geodesic distances on time ordered graphs. However, the authors assume the knowledge of all the nodes in the combined time steps beforehand which is not generally possible in dynamic networks, where both the edge and the node may appear or disappear in subsequent time steps. Further the time complexity for calculating temporal paths is comparable to static case, hence no real gain in terms of efficiency is achieved. There are also separate lines of work~\cite{rozenshtein2016temporal,taylor2015eigenvector,lerman2010centrality} that adapt the definition of eigenvector centrality and PageRank for dynamic networks; however, they require iterative parameter estimation and are not suitable for streaming setting. }
%\textcolor{red}{SB:how about papers on prediction ? I cut and paste these from section . Please check}

\noindent\textcolor{black}{\textbf{Social contact patterns}: In~\cite{kim2012centrality} the authors show that temporal human activity networks have periodic patterns and further devise prediction functions to calculate centrality of nodes at future time steps based on past history. Yang et. al.~\cite{yang2014predicting} develop a method, that combines concepts of preferential attachment and triadic closure to capture a node's prominence profile. They further show that the proposed node prominence profile method is an effective predictor of degree centrality. In~\cite{zhou2015temporal,zhou2017predicting} the authors observe that a node's temporal social contact patterns show strong correlation with its centrality. Based on this property they predict closeness centrality and apply their prediction to improve data forwarding in opportunistic mobile networks. One of the prime drawback of these previous works is in the datasets, which have been primarily restricted to human contact networks. To the best of our knowledge ours is the first work which applies centrality prediction on diverse real world and synthetic networks. We also use time series prediction models, which have not been employed for this task previously.}
%These key nodes play important role in information diffusion processes and epidemiology, due to their strategic position in the network. In case of static networks, influential nodes are estimated in terms of centrality metrics. The authors in~\cite{habiba2007betweenness} propose a generalization of betweenness centrality in dynamic frameworks using the idea of temporal shortest path. The authors show difference between traditional betweenness centrality and temporal betweenness centrality. On similar lines \cite{kim2012temporal,taylor2015eigenvector,tang2013applications} propose generalization of other node level centrality metrics in dynamic setting and motivate the need behind this objective. 

\noindent{\bf Time series forecasting}: Time series modeling have found a lot of application in economic analysis and financial forecasting ~\cite{hamilton1989new}. In the context of temporal networks major works incorporating time series formulation includes~\cite{scherrer2008description,hempel2011inner}. In \cite{sikdar2016time}, the authors leverage time series forecasting models to predict global properties of the network at a future time step. Although  our prediction scheme is based on a similar foundation, to the best of our knowledge ours is the first attempt towards identifying the top central nodes in a network.

\section{Definitions}
\subsection{Centrality measures}
\noindent{\em Closeness centrality} of a vertex $v$ is the average of the shortest distance between that node and all other nodes in the network.  It is calculated as $C_lC(v) = \frac{1}{\sum\limits_{s\neq v \in V} dis(v,s)}$, where $dis(v,s)$ is the length of a shortest path between $v$ and $s$. 

\noindent{\em Betweenness centrality} of a vertex $v$ is the ratio of the number of shortest paths between a vertex pair that passes through $v$ and all the shortest paths possible between that pair. It is given by  $BC(v) = \sum\limits_{s\neq v\neq t \in V} \frac{\sigma_{st}(v)}{\sigma_{st}}$, where $\sigma_{st}$ is the total number of shortest paths between $s$ and $t$, and $\sigma_{st}(v)$ is the total number of shortest paths between $s$ and $t$ that pass through $v$~\cite{freeman1978centrality}.

\subsection{Core periphery structure}
\label{theory}
Consider an undirected graph $G$ with $V$ as the set of vertices and $E$ as the set of edges. Let $K$ be a subset of vertices, i.e.,
$ K \subseteq V $ and $G(K)$ be the graph induced on $G$ by the vertices in $K$. $G(K)$ is considered to be $k$-core of the graph $G$ only if 

(i) For every $v \in K$, $d_{G(K)}(v) \geq k $ where $d_{G(K)}(v)$ denotes the degree of $v$ in $G(K)$;

(ii) For each $K \subset K' \subset V $ $\exists$ $u \in K' \backslash K $ such that $d_{G(K')}(u) < k $. All such $u$ form the $K'$ {\em{shell}} of the graph.

The latter condition enforces that there can be only one $K$-core in the graph $G$ where $K$ can vary from $1...d_{max}(G)$.
By this definition, the outermost core is numbered 1 and the core numbers increase consecutively until the innermost core numbered ${max}$.  For a connected graph, the entire graph is part of $1$-core. Repeated removal of nodes with degree one, leads us to $2$-core of the graph and so on. Note that we find the next highest core in the graph only if it exits.

The cores in the graph follow an inclusion relationship, i.e., $G(K_{max}) \subset G(K_{max-1}) \subseteq .....\subseteq G(K_{1})$. We define shells as the elements that are exclusive to a core, i.e., the shell $G(S_i)$ is the graph induced by the subset of vertices that are in $G(K_i)$ but not in $G(K_{i+1})$. The innermost core is its own shell, i.e., $G(S_{max})=G(K_{max})$. For ease of reading, we shall denote $G(S_{i})$ as $S_{i}$ and $G(K_{i})$ as $K_{i}$. \textcolor{black}{Further, to enhance readability, in Table~\ref{tab:notations} we note all the important symbols used in this paper.}

In this paper, we consider connected graphs, with the following properties. The innermost shell is connected \textcolor{black}{(empirically observed across all the networks in our dataset)}. The rest of the shells may or may not be connected and thus have multiple components. However, every component has at least one vertex that is connected to a higher numbered core. 

\begin{table}[!htb]
\centering
\begin{adjustbox}{max width=0.7\textwidth}
\huge {\textcolor{black}{
\begin{tabular}{ c  c }
\hline
{\bf Symbol } & {\bf  Description} \\
\hline 
$G(V,E)$  & graph $G$ with node set $V$ and edge set $E$ \\ 
$K$  & subset of nodes  \\ 
$G(K)$  & graph induced by nodes in $K$  \\ 
$d_{G(k)}(v) $   & degree of vertex $v$ s.t. $v$ exists in G \\
$K_{i}$ & set of nodes in the $i^{th}$ core \\
$K_{max}$ & set of nodes in the innermost core \\
$G_t$ & graph $G$ at time step $t$ \\
%$G^{data}_t$ & graph at time step $t$ for the dataset $data$ \\
$S_i$ & set of nodes in the $i^{th}$ shell \\
$S_{max}$ & set of nodes in the innermost shell \\
$P_{v \rightarrow u}^{X}$ & sequence of shortest paths between vertices $u$ and $v$ \\ 
$P_{v \rightarrow u}^{max}$ & sequence of shortest paths between vertices $u$ and $v$ with at least one node in $S_{max}$\\ 
$P_{v \rightarrow u}^{O}$ &  sequence of shortest paths between vertices $u$ and $v$ with no node in $S_{max}$\\
$P_{v \rightarrow u}=\infty$ &  no sequence exists connecting $u$ and $v$ \\
 \hline
 \end{tabular} }}
 \end{adjustbox} 
 \caption {\label{tab:notations}\label{notations} Table of notations.}
 \end{table}

\subsection{Our hypothesis}

We hypothesize that the high centrality vertices in many real world time-varying networks are more likely to be located in the innermost core. As a first step, we note that if most of the shortest paths pass through the innermost core, then the high centrality vertices would also be part of the innermost core. Moreover, if these vertices are tightly connected to each other, they can enhance each others' centrality values~\cite{ufimtsev2016understanding}. Thus, a dense innermost core through which most of the shortest paths pass provides us a smaller subset in which to search for high centrality vertices.

\begin{figure}
		\centering
        \begin{subfigure}[htb]{0.5\textwidth}
                \includegraphics[width=\linewidth]{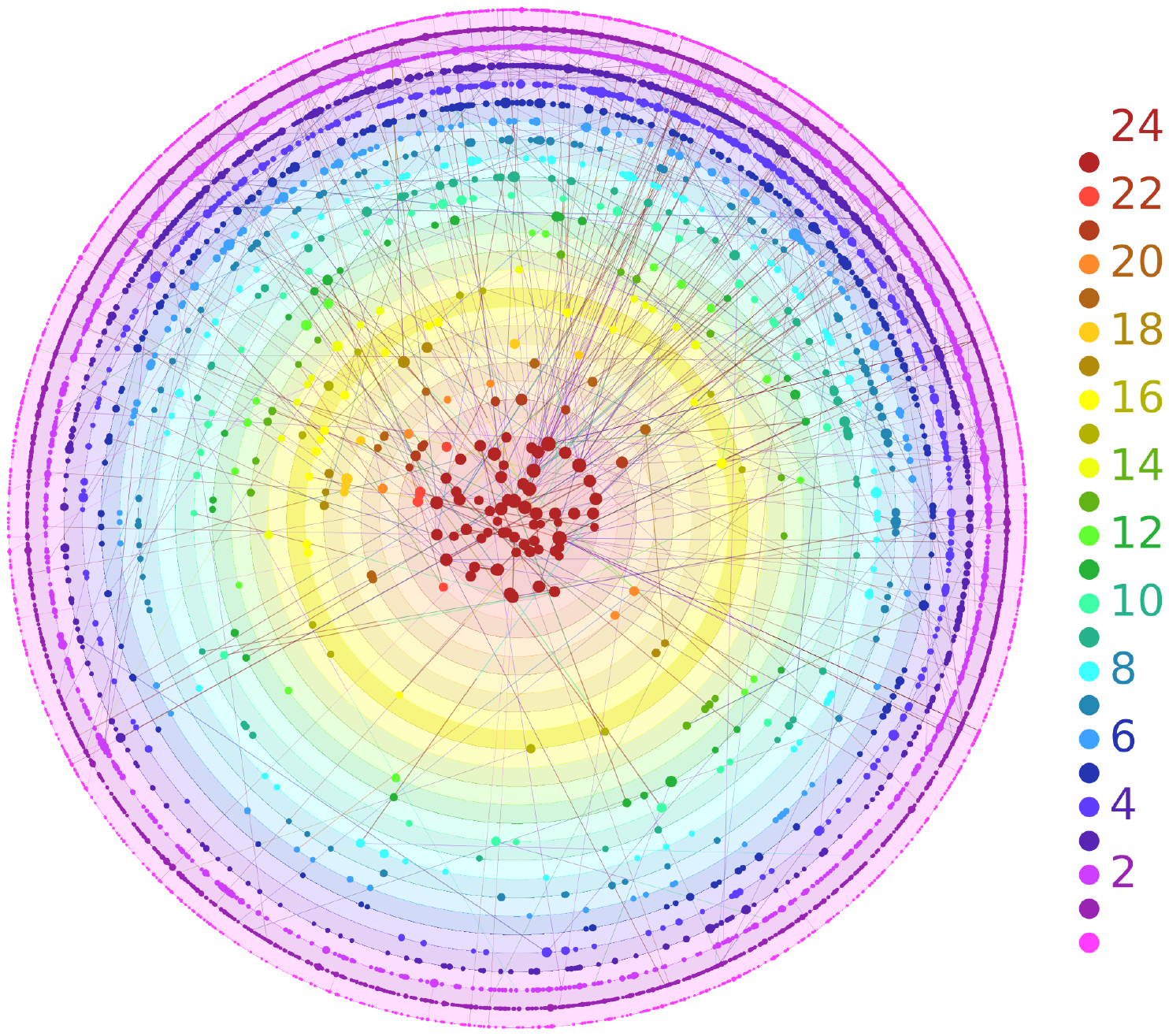}
                \caption{Caida network}
                \label{fig:As2}
        \end{subfigure}%
        \begin{subfigure}[htb]{0.5\textwidth}
                \includegraphics[width=\linewidth]{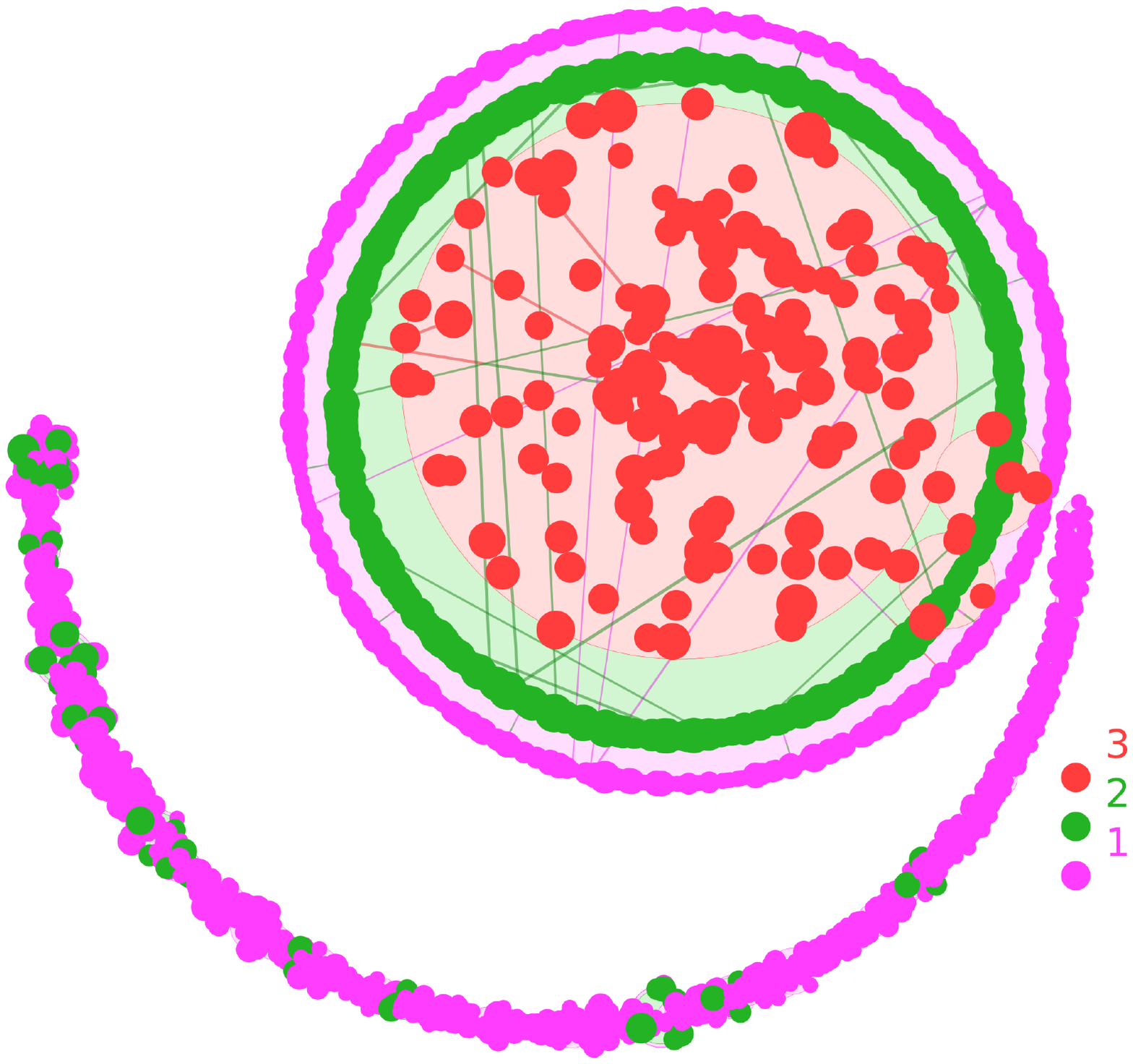}
                \caption{Facebook network}
                \label{fig:Fb}
        \end{subfigure}%
        \caption{\label{fig:Core}Visualization of the core-periphery structure with the corresponding shell index created using Lanetvi~\cite{alvarez2006lanet}; sizes of nodes are ordered based on the degree. Note that the Caida network has several layers of cores and and a small dense innermost core. In contrast, the Facebook network has only three cores of which the innermost core is sparse and predominant. Best viewed in color.}
\end{figure}

However, not all networks have such dense innermost cores. Figure~\ref{fig:Core} visually contrasts a network that has a small dense inner core (CA) to a network where the inner core is sparse and forms the bulk of the network (FW). From visual inspection, it is clear that it is easier to locate the high centrality nodes in the Caida network than in the network of Facebook users. 

Nevertheless, explicitly computing whether a network conforms to our hypothesis is expensive since that would lead to calculating the all pairs of shortest paths between the nodes repetitively as the network changes over time. We therefore provide a set of novel heuristics in the next section to classify networks based on the extent to which they conform to our hypothesis.
%\iffalse

%\fi

\section{Classification of the networks}
\label{sec:classification}

\begin{figure*}
\centering
\includegraphics[scale = 0.4]{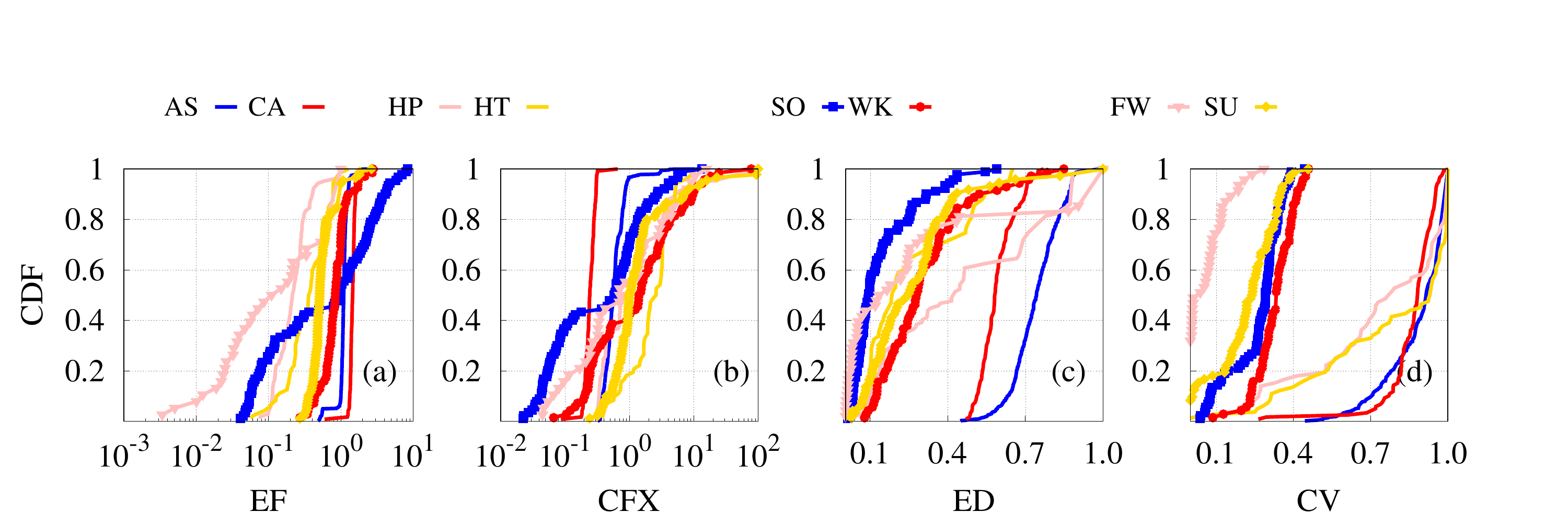}
\caption{\label{ccc_prop}{\bf Classification of the networks according the distribution of the parameters.} From left to right the parameters are, (a) fraction of inter-edges connected to the top core ($EF$), (b) average density of the non-top cores ($CFX$), (c) the density of the top core ($ED$) and (d) the overlap in the top-core at consecutive time steps ($CV$). Here AS, CA, HP, HT, SO, WK, FW and SU, given by the lines in different colors, represent the datasets. \textcolor{black}{The X-axis represents the time points and the Y-axis plots the Cumulative Distribution Function (CDF) for each of the parameters.}(\textcolor{black}{Please refer to Table~\ref{dataset} for detailed description.})}
\end{figure*}

In this section we propose the following four parameters to classify the temporal networks according to whether the high centrality vertices are within the innermost cores. Note that each of these parameters are less computationally expensive than computing the centrality of the vertices.  Our heuristics are as follows: 

\begin{itemize}
\item {\bf Fraction of inter-edges connected to the top core (EF)}: This metric is the ratio of the number of inter edges with one end point in top core to the total number of inter edges. {\em The higher the fraction, the more the network will have high centrality vertices in the top core.}

\item{\bf Average density of the non-top cores (CFX)}: This metric computes the average density of all cores, except the top one. The lower the density, the sparser the core, and the higher the average intra-core distance. The density of each core is computed as ratio of the number of intra-core edges in each core by the total possible edges between the vertices in the core. To find the average density we divide this value by the number of cores. {\em The lower this value the more the network will have high centrality vertices in the top core.} 

\item {\bf Density of the top-core (ED)}: We compute the density of the top-core which is the ratio of the number of intra-core edges in the top core to the total possible edges between the vertices in the core. {\em The higher this value, the more the network will have high centrality vertices in the top core.}

\item {\bf Top-core overlap (CV)}: This metric takes into account the changes in the top core structure over consecutive time steps. We measure the overlap as the Jaccard similarity between the vertices in the top cores of networks at time step $t-1$ and $t$. {\em The higher the value, the greater the overlap.}
\end{itemize}

\begin{figure}[!htb]
    \centering
    \begin{minipage}{\textwidth}
        \centering
   \includegraphics[width=0.85\linewidth, height=0.25\textheight]{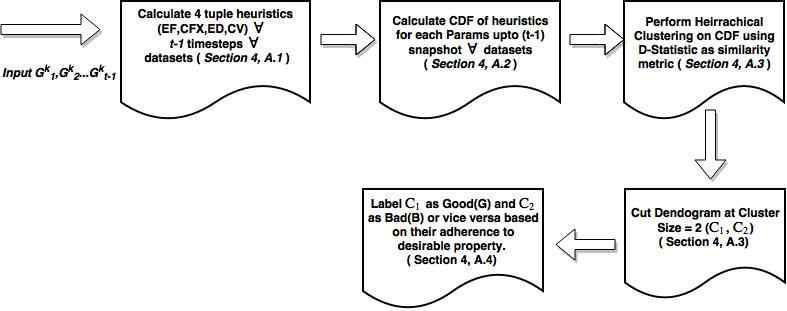}
    \end{minipage}%
    \caption{Classification framework.}
          \label{Fig:class}
\end{figure}

\textcolor{black}{We explain the classification framework based on the above parameters in the following steps.
\begin{itemize}
\item [(A)] For a set ($D$) of initial networks:
\begin{enumerate}
    \item We consider the $t-1$ temporal snapshots $G_1, G_2, ..., G_{t-1} \in D$. We calculate the 4 tuple heuristic for each snapshot. Hence each snapshot can now be represented as $(EF,CFX,ED,CV)$
    \item We calculate the cumulative distribution function (CDF) for each parameter from the $t-1$ values obtained for that parameter. For a particular parameter (say $EF$) we now have CDF for each of the initial datasets considered (figure~\ref{ccc_prop} shows the CDFs for the eight real world networks). Note that we compute the CDF over a range accumulated from all the parameter values obtained from all the previous time steps. In particular, it is not possible to compute the CDF if multiple values of the variable are not available, thus highlighting the use of multiple time points and consequently the dynamical properties of the network.
    \item We perform hierarchical clustering on each parameter, using $D$-statistic as the pairwise similarity measure between the CDFs of a particular parameter from two different networks. We cut the dendrogram at cluster size 2, hence having the clustering algorithm output 2 clusters ($C_1, C_2$) for each parameter. 
    \item If the mean value of parameter in $C_1$ conforms more to the desirable property, i.e., have high values for $EF, ED, CV$ or have low value for $CFX$, we label $C_1$ as good ($G$) and $C_2$ as bad ($B$) or vice versa. We have outlined our framework in figure~\ref{Fig:class}.
\end{enumerate}
\item [(B)] For each new (unseen) network:
\begin{enumerate}
\item Once again we compute the CDFs for each of the four parameters from the $t-1$ snapshots.
\item Next we obtain the similarity ($D$-statistic) of the CDF of a particular parameter with the centroid of both $C_1$ and $C_2$ (i.e., the traditional Rocchio technique~\cite{baeza1999modern}). 
\item Finally, we classify the network to that class to which it is more similar (based on the similarity with the centroid of the class.) 
\end{enumerate}
\end{itemize}
We use eight real world networks as the initial set and 20 synthetic networks as unseen and classify them based on the Rocchio scheme. The category for each network are presented in Table~\ref{tab:res} and Table~\ref{new:results} for real and synthetic networks respectively. For all the parameters, except $CFX$, higher value is good.  For $CFX$, lower value is desirable. This classification matches with the results in Table~\ref{tab:res} and Table~\ref{new:results}, i.e., if $EF$, $ED$ and $CV$ are high and $CFX$ is low then the high centrality vertices in the networks can be predicted with higher accuracy using our scheme.}

\iffalse{}
These four values are computed over each time steps (or consecutive time steps for CV). We compute the cumulative distribution function (CFD) for the series of values for each parameter and for each network . We then cluster the distribution functions for each network, by performing hierarchical clustering with the similarity value between any two distributions as the $D$-statistic value calculated using Kolmogorov-Smirnov test (this essentially determines how close the two distributions are, with lower value indicating higher similarity). For each parameter we place the networks in two distinct classes - Good (G) and Bad (B), based on how well they conform to the desired value of the parameter. Each network is hence categorized by a four tuple $(EF,CFX,ED,CV)$. The category for each network are presented in Table~\ref{tab:res}. For all parameters, except CFX, higher value is good.  For CFX, lower value is desirable. This classification matches with the results in Table~\ref{tab:res}, if EF, ED and CV are high and CFX is low then the high centrality vertices in the networks can be predicted with higher accuracy. 
\fi

Figure~\ref{ccc_prop} shows the distribution of the different parameters for each of the real world networks. The parameters for top core overlap and top core density form the most distinct clusters. The clusters for fraction of edges to the top core, and intra-core density are not as distinct, indicating that almost all the networks conform to our hypothesis at least to some extent and that the size and density of the top core are the critical factors in determining predictability. The groupings for each parameter and for each network are noted in Table~\ref{tab:res} (real networks) and Table~\ref{new:results} (synthetic networks).

\section{Algorithm for predicting high centrality vertices}\label{algorithm}
\label{prediction}
%\noindent
\begin{algorithm}
%\DontPrintSemicolon
\KwData{$G_{t-1}(V,E)$ , graph at time step $t-1$; $l_k$ set of top core vertices at time $(t-2)$} %\TODO{AM: Is it $(t-1)$ of $(t_k-1)$}}
\KwResult{EF, CFX, ED, CV}
\SetKwFunction{FindCore}{FindCore}
\SetKwFunction{Jaccard}{Jaccard}
\SetKwFunction{MaxCore}{MaxCore}
\SetKwFunction{Centrality}{Centrality}
\CommentSty{\color{black}}
$c[v] \leftarrow $ \FindCore{G} \tcp*[h]{Returns core number of each vertex $v\in G$} \;
$C_m \leftarrow  \MaxCore{c}$ \tcp*[h]{Returns max core number from c} \; %\TODO{AM: $C$ Undefined.}  
$L_k \leftarrow \{ \}$ \;
$N(v) \gets $ neighbors of $v$ \;
$tempV[C_m],tempE[C_m], tempD[C_m] \leftarrow [0]$ \;
\For { $\forall v \in V $} {
 \For{$u \in N(v)$ } {  
 \If {$c[u] \neq c[v]$} 
 {
  \If {$(c[u] == C_m \lor c[v] == C_m)$} {
          $e_m \leftarrow e_m+1$ \;
  		}
    $ e_i \leftarrow e_i+1 $ \;
  }
}
}

$EF \leftarrow e_m/e_i $ \;

\For { $\forall v \in V $} {
 \For{$u \in N(v)$ } {
   \If {$c[u] == c[v] $} {
     $tempE[c[u]] \leftarrow tempE[c[u]]+1 $\;
     $tempV[c[u]] \leftarrow tempV[c[u]]+2 $\;
     }
 }
 }

 \For {$ u < C_m $} {
    $tempD[u] = 2*tempE[u]/(tempV[u]*(tempV[u]-1))$ \;
 }
 \For{$u < C_m-1$}{
        $d \leftarrow d+tempD[u]$ \;
 }
 $CFX \leftarrow d/(C_m-1)$ ; 
 $ED \leftarrow tempD[C_m] $ \; 
 \For { $\forall v \in V $} {
   \If {$c[v] == C_m$} {
   		$L_k \gets L_k \cup v$\;
   }
 }
$ CV \leftarrow $ \Jaccard{$l_k,L_k$}\;
\Return{EF, CFX, ED, CV}\;
\caption{Calculate the classification parameters.}
\label{algo:CCParams}
\end{algorithm}

\begin{algorithm}
%\DontPrintSemicolon
\KwIn{$G_{t}(V,E)$ , graph at time step $t$, $N_v = |V|$  }
\KwOut {T[k]}
\SetKwFunction{FindCore}{FindCore}
\SetKwFunction{MaxCore}{MaxCore}
\SetKwFunction{MDVid}{MDVid}
\CommentSty{\color{black}}
$temp[N_v] \leftarrow [0]$\; 
$m,j \leftarrow 0$\;
%$m \leftarrow 0$\; 
$c[v] \leftarrow $ \FindCore{G}\; 
$C_m \leftarrow $ \MaxCore{C} \tcp*[h]{Returns max core number from c}\; 
\For {$i \leftarrow 1 $ \KwTo $ N_v $}
 {
    \If { $ c[i] == C_m $ }
    {
      $temp[j] \leftarrow i$\;
      $j \gets j+1$\;
    }
 }
\For{$i \gets 1$ \KwTo $k$} {
   $n \gets$ \MDVid{temp} \tcp*[h]{Max Degree vertex id}\;  
   T[m++] $\leftarrow  temp[n]$ \;
   $temp[n] \leftarrow 0$\; 
}
\Return{$T$}\;
\caption{Predict top central vertices}
\label{algo:predict}
\end{algorithm}

Our prediction framework is composed of three steps. In the first step we classify the network based on the four parameters introduced in the previous section. In the second step we use the already computed overlap time series to predict the future overlap values using ARIMA models. In the final step we identify high centrality vertices in the network at the next time step. %newly arrived graph. 

\begin{itemize}
\item \textcolor{black}{{\bf Step 1: Classifying the networks based on the parameter values across time steps}:}

\textcolor{black}{Our first step is to classify the networks to see whether the vertices with high centrality are consistently located in the top core over all the $t$ time steps. To obtain this classification, we compute the parameters, defined in section~\ref{sec:classification}, for networks at every time step from $1$ to $t$. The complexity for computing the parameters are as follows:} 

\textcolor{black}{To compute these parameters, we first need to compute the core numbers of the vertices. We do this using the function  $FindCore$ (see line 1, Algorithm~\ref{algo:CCParams}), which  implements the algorithm presented in~\cite{batagelj2003m} and has a complexity of $O(|E|)$, where $|E|$ is the number of edges.}

\textcolor{black} { For computing fraction of inter core edges connected to the top core, i.e., {\em{EF}}, (lines 6--12, Algorithm~\ref{algo:CCParams}) , we need to iterate over all edges once and keep count of the number of edges where the end points belong to different cores and at least one of them is part of the top core. Thus the complexity is $O(|E|)$.} 

\textcolor{black}{To compute the density of each core, we sum the number of the edges whose endpoints are both in that core and then divide this sum by the total number of possible edges in the core. Using these values we can compute the average density of all the non-top cores, i.e., {\em{CFX}}, as well as the density of top core, i.e., {\em{ED}}, (lines 13-22). Once again computing these parameters requires us to go through all the edges, and thus has complexity $O(|E|)$.} 

%Intra core density of all the cores in the graph can be computed in $O(|E|)$ (here $|E|$ is the number of edges) by a single traversal of the edges of the graph and keeping a count of the number of internal edges at each core normalized by all possible edges. Average density of non top cores referred to as {\em{CFX}} and the density of top core ({\em{ED}}) can be easily extracted from the pool of all core densities (lines 10-18).  

\textcolor{black}{The overlap between the top core vertices, {\em CV}, in the networks over two consecutive time steps is computed using Jaccard similarity (see Algorithm~\ref{algo:CCParams}). Finding the vertices in the top core, requires us to go over all the vertices and identify their core numbers (complexity $O|E|$). The complexity of computing the Jaccard similarity is linear to the number of vertices in the top core.}

 \textcolor{black}{Once we obtain these parameters for all the networks over the time steps $1$ to $t$, we classify each network by the technique described in the previous section. Based on the classification, the network might fall in the good class $G$ (i.e., majority or at least equal number of parameters fall in the good class) or in the bad class $B$. {\em Note that this classification, requires information from not just one time step, but from multiple time steps thus establishing how our framework is dependent on the dynamical properties of the network.} If the network falls in class $G$ we proceed to Steps 2 and 3.}

\item \textcolor{black}{{\bf Step 2: Estimating overlap among the top central vertices}: After analyzing the networks from time steps $1$ to $t$, consider the network $G_{{t+1}}$ at time step ${t+1}$. Even if the network is not available, we leverage information about the Jaccard overlap between the top core vertices for consecutive time steps from $1$ to ${t}$, and use the autoregressive-integrated-moving-average (ARIMA) algorithm~\cite{box2015time} to estimate the  overlap in top cores of $G_{t}$ and $G_{{t+1}}$.}

\item \textcolor{black}{{\bf Step 3: Identifying the top central vertices}: If the network is classifed into class $G$ (good), then we identify the top central vertices in the new network $G_{{t+1}}$, using Algorithm~\ref{algo:predict}.} 

\textcolor{black}{ We identify the vertices in the top core of the new graph using the $FindCore$ function (line 3, Algorithm~\ref{algo:predict}). Then we implement a $m$-search algorithm to extract the $m$ highest degree vertices (lines 5-12, Algorithm~\ref{algo:predict}) of the graph, that are in the top core. These vertices are marked to be the high centrality vertices.}
\end{itemize}

\begin{figure}[!htb]
	\vspace{-0.6cm}
    \centering
    \begin{minipage}{\textwidth}
        \centering      \includegraphics[width=0.85\linewidth, height=0.3\textheight]{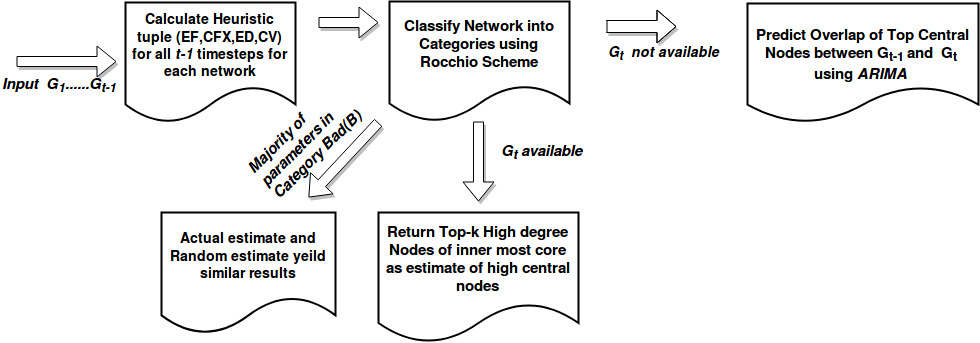}
    \end{minipage}%
    \caption{Prediction framework.}
    \label{Fig:Predicts}
\end{figure}

\textcolor{black}{If the network falls in the class \textit{B}, a random selection of nodes from the network and the actual high central nodes perform equivalently with respect to path based centrality (substantiated by the validation results in section~\ref{valid}.)}

We now provide further details on the most crucial steps 2 and 3 in the rest of this section.

\subsection{Estimating the extent of overlap}
For a given temporal network ($G$) with network snapshots $G_{1}, G_{2}, \ldots$ observed at times $1,2, \ldots$, we calculate for initial few time steps the top (5) 10 central vertices based on betweenness and closeness. The overlap $l_{c}^{t}$ (closeness) and $l_{b}^{t}$ (betweenness) is calculated as the Jaccard overlap between the top (5) 10 central vertices between network snapshots $G_{t}$ and $G_{t+1}$. Note that this is calculated for the initial few time steps to eliminate the cold start problem. We next represent $l_{c}^{t}$ ($l_{b}^{t}$) as a set of points ordered in time or equivalently a time series. This representation allows us to leverage time series forecast models to predict the values of $l_{c}$ and $l_{b}$ at a future time step. Specifically, we use the ARIMA model to fit the resulting time series. On fitting an ARIMA($p$,$d$,$q$) model to a time series we obtain an auto-regressive 
equation of the form -
\begin{center}
 $y_{t}=\alpha_{1}y_{t-1}+\cdots+\alpha_{p}y_{t-p}+\beta_{1}e_{t-1}+\cdots+\beta_{q}e_{t-q}+c$
\end{center}
where $y_{t}$ represents the value of the time series at time $t$, $e_{t},e_{t-1}\ldots$ are the white noise terms and $\alpha_{i}$ and $\beta_{i}$ are parameters of the model. To summarize, given the centrality overlap values till $t-1$, we are able to predict it at time $t$. If the top central vertices are known for $G_{t-1}$ and we know the extent of overlap among top central vertices between $G_{t-1}$ and $G_{t}$ (which we predict using our proposed method), we can roughly estimate the top central vertices in $G_{t}$ given the predicted overlap value is high and the prediction error is low. Note that {\em for the above technique to work, the network $G_t$ itself is not required}. We show later (section~\ref{emp_res}) that our method is indeed able to predict the overlap values for certain networks with very low error. 

\subsection{Identifying the top central vertex} \label{subsec1}
The first step of the proposed prediction scheme allows for estimating the extent of overlap among the top central vertices between the two consecutive snapshots of the network.
%and, thereby, roughly estimates the top central vertices without knowing the structure of the network. 
In this step we further refine the prediction when the network $G_{t}$ is available. Specifically, instead of explicitly calculating the centrality values, we use the algorithm described in \cite{batagelj2003m} to obtain the core-periphery structure and, thereby, identify the vertices in the top core. Our analysis in section~\ref{proof_sketch}, suggests that these vertices are the most likely candidates for being the high central ones in the network. To obtain the top $m$ central vertices in the system, we rank the vertices in the top core based on their degree and filter out the top $m$ vertices (see Algorithm~\ref{algo:predict}). Note that this step has complexity $O(|E|)$ (for computing the core and ranking the top-core vertices by degree), which is significantly less than $O(|V|*|E|)$, the complexity for computing all the closeness and betweenness centrality vertices and then ranking them.

\textcolor{black}{The innermost core hence acts as a `container' for the high central nodes in certain class of networks which we explain in Section~\ref{proof_sketch}. We also empirically show this by performing the following experiment: instead of predicting the high degree nodes from within the top core as the high central nodes we directly predict the globally highest degree nodes agnostic of which core they belong to. This results in drastically poorer $F1$-score values as noted in Table~\ref{tab:r1}.}

\textcolor{black}{We also compare the results obtained from our prediction algorithm against existing algorithms for forecasting high centrality nodes in temporal networks~\cite{kim2012centrality}. Note that all these schemes are compute intensive as they require explicit computation of the centrality values in all the earlier time steps for prediction in the current time step unlike our approach. For both closeness and betweenness centrality our results are superior compared to existing state-of-the-art in almost all cases of the cases.} 

\textcolor{black}{We illustrate the complete flowchart of our prediction framework in figure~\ref{Fig:Predicts}. Note that all the subsequent steps of prediction are dependent of the initial classification step which is completely regulated by the temporal dynamics of the network under consideration.}

In section~\ref{emp_res}, we show that our proposed method is indeed able to identify a large fraction of central vertices in the network without explicitly calculating the centrality values (betweenness and closeness) for each vertex. 
 \medskip

\section{Experimental setup}
In this section we describe the networks in our test suite. \textcolor{black}{We consider 8 diverse type of real world networks. We also consider 20 synthetic networks of varying numbers of nodes, edges and temporal snapshots and generated using two different tools -- Musketeer~\cite{gutfraind2015multiscale} and Dancer~\cite{benyahia2016dancer}.} 

\subsection{Test suite of real-world networks}
We consider a diverse set of benchmark networks of various sizes and over discrete timescales. The networks are collected from public repositories made available at~\cite{snapnets,kunegis2013konect}. Our networks can be grouped into three different categories according to the application domain. Given below is a brief description of the categories, the networks in them, and how we obtained the time series network from the data. The sizes of  the networks is given in Table~\ref{dataset}.

\begin{itemize}
\item{\bf Autonomous systems network:} The Internet is sub-structured as interconnected subgraphs of highly connected routers. These subgraphs are known as autonomous systems (AS). Each AS exchanges traffic with neighbors (peers) using BGP (Border Gateway Protocol). Here we use two example networks; both were created using BGP table snapshots and made publicly available. 

\noindent{\em AS}: The first dataset (AS)  was collected from University of Oregon Route Views Project and it contains 733 daily data traces between autonomous systems which span an interval of 785 days from November 8 1997 to January 2 2000. 

\noindent{\em CA}: The second dataset was collected by Center for Applied Internet Data Analysis (CA)  from January 2004 to November 2007 and it comprises anonymized interaction of ISP's.

\item {\bf Citation network:} This type of network connects two papers if one paper cited the other. Although the links are directed, for purpose of our experiments we consider them to be undirected. 

We use citation data from two different research topics in high energy physics. For both these networks, every paper is timestamped by the submission time to the archive. We also have a list of papers which are cited by a submitted publication. Based on this information, we created an aggregated growing network in terms of months and considered each network as a distinct snapshot. For citation network a node once added is not deleted. This is not the case for the other classes of networks.

\noindent{\em HepPh} ($HP$): (High Energy Physics Phenomenology) is a citation graph from the e-print arXiv and covers all citations within a dataset of 34,546 papers with 421,578 edges. Citation between paper $i$ and $j$, is represented as an edge. The data covers an almost complete set of papers from January 1993 to April 2003. 

\noindent{\em HepTh} ($HT$): (High Energy Physics Theory) is a citation graph is similar to HP, with 27,770 papers with 352,807 edges. The data focuses on papers from January 1993 to April 2003.

\item{\bf Social communication networks:} These networks are of interactions via different types of  social media. We study four different networks with each individual edge accompanied by unique timestamp. For all these networks we aggregated all the edges appearing in the same month and created a single temporal snapshot of the network per month. 

\noindent{\em StackOverflow} ($SO$): On stack exchange web sites, users post questions
and receive answers from other users, and users may comment on both questions and answers. A temporal network is derived by creating an edge $(u, v, t)$ if, at time $t$, user $u$: (1) posts an answer to user $v$'s question, (2) comments on user $v$'s question, or (3) comments on user $v$'s answer.

\noindent{\em Facebook Wall} ($FW$): The edges of this dataset are wall posts between users on Facebook located in the New Orleans region. Two users are connected if they post on the same wall.

\noindent{\em Wiki Talk} ($WT$): This dataset represents edits on user talk pages on Wikipedia. An edge $(u, v, t)$ signifies that user $u$ edited user $v$'s talk page at time $t$.

\noindent{\em Superuser} ($SU$): This dataset is derived from question answer site Superuser which exists for computer enthusiasts. As in the case of  StackOverflow, an edge $(u, v, t)$ exists if, at time $t$, user $u$: (1) posts an answer to user $v$'s question, (2) comments on user $v$'s question, or (3) comments on user $v$'s answer.

\subsection{Test suite of synthetic networks}
\textcolor{black}{We also consider 20 synthetic networks generated using the multi-scale network generation tool Musketeer~\cite{gutfraind2015multiscale} and the dynamic attributed networks generation tool Dancer~\cite{benyahia2016dancer}. We generate several temporal snapshots of each network with various core structures to present additional prediction results and, thereby, further strengthen our hypothesis.} 
\end{itemize}

\begin{table}
\centering
\begin{adjustbox}{max width=0.6\textwidth}
\huge {
\begin{tabular}{ c  c  c  c  c }
\hline
{\bf Network } & {\bf Nodes} & { \bf Unique Edges} & {\bf Temporal Edges} & {\bf Time Span}  \\ \hline
AS & 7716 & 27183 & 57,05405 & 732 \\ 
CA & 31255 & 111564 & 54,85410 & 120 \\ 
HP & 34564 & 4,21578 & 4,21578 & 124 \\ 
HT & 27770 & 3,52807 & 3,52807 & 124\\ 
FW & 46952 & 274,086 & 876,993 & 48 \\ 
SU & 194,085 & 9,24886 &  1,443,339 &  92 \\ 
WT & 1,140,149 & 3,309,592  & 7,833,140   & 73 \\ 
SO & 2,601,977 &  36,233,450 & 63,497,050 & 92 \\ \hline
 \end{tabular} }
 \end{adjustbox} 
 \caption {\label{dataset} Test suite of real world networks used for our experiments. For AS and CA the time span is measured in days, for all others in months. Combined entire edge stream for any dataset comprises the temporal edges; unique edges are temporal edges with duplicates removed.}
 \vspace{-6mm}
 \end{table}

\section{Empirical results}
\label{emp_res}
In this section we present the empirical results to demonstrate the effectiveness of our prediction algorithms as proposed in section~\ref{prediction}. 
\subsection{Results on real world networks}
Our prediction scheme consists of two steps and we evaluate each of them separately.
\subsubsection{Extent of overlap} We measure the effectiveness of the prediction scheme using cross-validation technique. More specifically we consider each network and predict the overlap at different time steps for both betweenness and closeness. If for a given time step $t$, the original overlap value is $o_t$ and the predicted value is $p_t$, we define the error in prediction $error_{t}$ as \vspace{-1mm}
\begin{center}
 {\large$error_{t}=\frac{|o_{t}-p_{t}|}{o_{t}}*100$}
\end{center}
\vspace{-1mm}
Note that for this prediction we assume that the actual overlap values for a certain time stretch till $t-1$ is available. In our experiments this time stretch has been set to 20 time steps until $t-1$\footnote{We tried with other stretches of size 15, 25 etc. The results do not seem to be affected by such minor variations. Ideally this size should not be too large thus consuming a lot of data for prediction, nor it should be too small thus having too few points to correctly predict. Through experimentation, we find that a size close to 20 strikes an ideal balance.}. 

In Table~\ref{tab:res} we present, \textcolor{black}{along with the mean original overlap values}, the average and standard deviation of the error percentage across all the time steps for each dataset for both betweenness and closeness centrality. We present results considering top 10 high closeness and betweenness centrality vertices. However, our scheme also works very well for even a much more restricted set of even five vertices (see Table~\ref{tab:res})\footnote{Note that if we keep increasing the number of top vertices, the prediction results can only get better. Through experiments, we observe that small numbers like 5 and 10 are judicial choices.}. We observe that the error is very low for the networks in $GGGG$ (all parameters good) category (AS, CA) while those in $BBBB$ (all parameters bad) category (WT, FW, SU) the error rates are much higher. \\
\noindent{\bf Comparison with other time series models}: We further look into other models of time series 
prediction (AR, MA and ARMA in specific) and mean prediction error (both betweenness and closeness) for these models are reported in Table~\ref{tab:base}. For networks in $GGGG$ category, these simpler models are able to predict the extent of overlap with very low error but their 
efficiency reduces as we move toward the other classes of networks.

\begin{table}[htpb] 
\centering
\caption {\label{tab:res} Classification as well as the prediction performance for the datasets used for evaluation. Each dataset is classified as a four tuple $(EF,CFX,ED,CV)$ (column 1) with $G$ representing good and $B$ representing bad. Mean ($\mu$), std. dev. ($\sigma$) are reported for both prediction error and $F1$-score (columns 3 to 8). The categories are colored as per the groups they belong. Note that the higher the number of $G$s in the category, the more accurate the prediction results.}
\begin{adjustbox}{width=\textwidth,  height=1.8cm}
\Huge { %
\begin{tabular}{c c c  c  c  c  c  c  c }
\hline
{\bf N/w Category} & {\bf N/w Name} & { \bf \textcolor{black}{Mean overlap}}  &
\begin{tabular}[c]{@{ }c@{ }}
{\bf Close. Pred.}\\
{ \bf (top 5, $\mu$, $\sigma$)}
\end{tabular} & \begin{tabular}[c]{@{ }c@{ }}
{\bf Bets. Pred.}\\
{ \bf (top 5, $\mu$, $\sigma$)}
\end{tabular} & \begin{tabular}[c]{@{ }c@{ }}
{\bf Close. Pred.}\\
{ \bf (top 10, $\mu$, $\sigma$)}
\end{tabular} &  \begin{tabular}[c]{@{ }c@{ }}
{\bf $F1$-score}\\
{ \bf (top 10, $\mu$, $\sigma$)}
\end{tabular}  & \begin{tabular}[c]{@{ }c@{ }}
{\bf Bets. Pred.}\\
{ \bf (top 10, $\mu$, $\sigma$)}
\end{tabular} & \begin{tabular}[c]{@{ }c@{ }}
{\bf $F1$-score}\\
{ \bf (top 10, $\mu$, $\sigma$)}
\end{tabular}  \\
   \hline
%\rowcolor{LightCyan}
$GGGG$& AS &  0.79 &.62,10.61 & 7.78,10.86 & 5.69,6.37 & 0.81,0.06  & 6.97,7.68 & 0.72,0.08  \\ 
%\rowcolor{LightCyan}
$GGGG$& CA & 0.82 &11.68,26.82 & 3.59,5.67 & 8.76,6.02 & 0.77,0.08 & 9.17,6.47 & 0.64,0.07\\ 
\hline
%\rowcolor{Gray}
$GGBG$& HT & 0.56 &12.52,11.71 & 20.79,16.10 & 26.96,17.44 & 0.42,0.35 & 20.74,14.86& 0.52,0.30 \\ 
%\rowcolor{Gray}
$GGBG$& HP & 0.43 &26.76,17.44  & 20.74,14.86  & 11.64,5.76 & 0.42,0.33 & 14.22,11.23 & 0.46,0.29\\
\hline
%\rowcolor{ofwhite}
$BBBG$& SO& 0.29 &34.92,33.39 & 45.30,38.48 & 27.96,21.69 & 0.35,0.26 & 26.15,24.72 & 0.39,0.30 \\ 

$BBBB$& WT & 0.05 &47.10,36.56 & 59.84,43.36 & 41.77,31.33& 0.32,0.17 &36.55,28.70& 0.31, 0.22\\ 
$BBBB$& FW& 0.09 &131.89,148.42 & 169.57,158.51 & 109.90, 92.39& 0.24,0.25 & 56.19, 34.95& 0.20,0.19\\ 

$BBBB$& SU& 0.03 &44.45,40.14 &167.25,130.42 & 147.06,106.53 & 0.02,0.09 & 32.58,40.14& 0.18,0.21\\ \hline
 \end{tabular}}%
 \end{adjustbox} 
 \end{table}

\begin{table}[]
\centering
\caption{Prediction performance of AR, MA and ARMA time-series prediction models across all the datasets. Both mean ($\mu$) and std. dev. ($\sigma$) are reported in each case. Predictions are made considering top 10 central vertices.}
\label{tab:base}
\begin{adjustbox}{max width=0.85\textwidth,height=1.4cm}
%\end{}
\begin{tabular}{c c c c c  c c}
\hline 
         & \multicolumn{2}{c}{AR}     & \multicolumn{2}{c}{MA}     & \multicolumn{2}{c}{ARMA}   \\ \hline
Datasets & $BC(\mu,\sigma)$           & $C^lC(\mu,\sigma)$           & $BC(\mu,\sigma)$           & $C^lC(\mu,\sigma)$           & $BC(\mu,\sigma)$           & $C^lC(\mu,\sigma)$                     \\ \hline 
%\rowcolor{LightCyan}
AS       & 7.72, 10.75  & 6.93, 12.37  & 7.58, 10.01  & 6.37, 9.53   & 7.48, 10.16  & 6.45, 10.37     \\
%\rowcolor{LightCyan}
CA       & 10.78, 8.82  & 9.26, 6.13   & 10.45, 9.11  & 9.15, 6.17   & 10.56, 8.97  & 9.23, 6.12      \\ 
%\rowcolor{Gray}
HT       & 21.72, 12.65 & 27.58, 11.09 & 22.15, 11.71 & 28.56, 19.21 & 22.67, 14.63 & 26.82, 14.32    \\ 
%\rowcolor{Gray}
HP       & 19.73, 14.97 & 26.96, 17.44 & 21.31, 15.29 & 24.37, 13.35 & 20.41, 14.98 & 25.56, 18.14    \\
%\rowcolor{ofwhite}
SO       & 35.02, 33.30 & 33.62, 32.46 & 34.92, 33.39 & 32.36, 31.29 & 34.56, 32.42 & 32.12, 32.62    \\
WT       & 62.09, 43.81             & 45.86, 35.98             &  60.54, 45.79            &   44.63, 35.64           & 60.55, 44.39             &   44. 56, 34.88              \\
FW       &  135.48, 122.68            & 241.68, 215.57             & 132.58, 123.45             & 163.68, 149.42             & 132.26, 122.83             &  157.70, 150.34              \\ 
SU       & 44.59, 43.49             &   167.25, 145.42           &  41.16, 38.23            & 169.26, 143.45             &  45.97, 44.42            &   166.52, 141.78             \\ \hline
\end{tabular}
\end{adjustbox}
\end{table}

\noindent{\bf Prediction using previously predicted values}: Note that in the experiments discussed above, at each point of prediction ($t$) we consider the original values of the series between $t-1$ and $t-20$. We further explore the case where instead of using the original values we use the predicted values. More precisely, for predicting the values at point $t$, we use the predicted values between $t-1$ and $t-20$. Note that to avoid cold start problem we use the original values for the first point of prediction but as we move along we keep on using the predicted values instead of the original ones. 

Consequently, we observe that the prediction error increases drastically in case of FW and SU while for AS and CA, the effect is negligible (see Table~\ref{tab:rec}). We report the results considering the top 10 vertices; however, the results considering five vertices show exactly similar trend. This once again demonstrates that  our classification can identify networks where high centrality vertices can be predicted accurately.   

\begin{table}[!htp]
\centering
\caption{Percentage error in prediction (mean$(\mu)$, std. dev. $(\sigma)$) for all the datasets using predicted values repeatedly for prediction. The results are reported considering top 10 central vertices. }
\label{tab:rec}
%\scalebox{0.8}{
\begin{adjustbox}{max width=\textwidth,max height=1.4cm}
\begin{tabular}{ccccccccc}
\hline
Dataset & AS$(\mu,\sigma)$ & CA$(\mu,\sigma)$ & HP$(\mu,\sigma)$ & HT$(\mu,\sigma)$ & SO$(\mu,\sigma)$ & WT$(\mu,\sigma)$ & FW$(\mu,\sigma)$ & SU $(\mu,\sigma)$\\ \hline
CC      & 6.30,7.37   & 13.02,11.25   & 14.56,10.21   & 32.46,30.77   &  37.58,35.07  & 76.40,91.85    & 158.86,121.21   &  172.16,114.23  \\
BC      & 8.21,13.30   & 10.92,16.12 & 25.91,14.29   & 19.54,12.48   & 32.98,25.45   & 44.42,28.57   & 73.66,30.95   & 41.02,26.27   \\ \hline
\end{tabular}%}
\end{adjustbox}
\end{table}

\subsubsection{Identifying top central vertices} Once the network has arrived in time, we further refine our prediction to identify exactly the top 10 central vertices based on betweenness and closeness centrality values. We further obtain the predicted set of top central vertices based on the prediction scheme described in section~\ref{subsec1}. To measure the effectiveness of our scheme we compute the $F1$-score between the predicted and the original set. We repeat the result for each time step and the average $F1$-score as well as its standard deviation are reported in Table~\ref{tab:res}. The results are separately reported for betweenness and closeness centrality. We again observe that the best prediction result is obtained for the networks in the $GGGG$ category. For the networks in the $BBBB$ category the obtained $F1$-score is the least. Once again, while we report results considering the top 10 vertices, the results for the top five vertices show an exactly similar trend.

\subsection{ Necessity of computing the core}
\label{subsec:tab5}
\noindent\textcolor{black}{At this point one might question whether or not computing the innermost core is indeed required. We posit that the innermost core acts as a `container' for the high central nodes in the predominantly $G$ class of networks. In order to show the utility of the innermost core computation, we perform the following experiment: instead of predicting the high degree nodes from within the innermost core as the high central nodes we directly predict the globally highest degree nodes agnostic of which core they belong to. This results in drastically poorer $F1$-score values as noted in Table~\ref{tab:r1}}.

\subsection{Comparison with baselines}
\label{subsec:bseline}
\textcolor{black}{In this section we compare our method with three existing baselines taken from ~\cite{kim2012centrality}. Here the authors estimate the centrality scores of nodes at a future time step $t$ using $r$ previous centrality scores. However, these baselines require that all the nodes be present at all time points. Note that this is not the case for our datasets. Hence to give full advantage to the baseline models, we choose $r>>1$ so that we can obtain a non-zero average centrality value for each node in the network (even though it does not appear at all time points). For readability we briefly describe the three baselines below. For the first baseline -- Uniform -- the authors estimate centrality of a node at time step $t$ by taking an uniform average of the node's centrality in $r$ previous time steps. In the second (W1) and third (W2) baselines the centrality of a node at time step $t$ are calculated as weighted averages of the $r$ time steps. For W1, the weights of the previous $r$ centrality values go as $(\frac{1}{d})$ where $d$ is the distance of the prediction point $t$ from the previous time point being considered. For W2, the weights of the previous $r$ centrality values go as $(\frac{1}{\sqrt{d}})$ where $d$ is again the distance of the prediction point $t$ from the previous time point being considered. Note that the maximum value of $d$ is $r$. Note that all these schemes are very compute intensive, unlike our approach, as they require explicit computation of the centrality values in all the $r$ earlier time steps for prediction in the current time step $t$. In Tables~\ref{tab:r1} and~\ref{tab:r2} we compare the $F1$-scores of our prediction algorithm with the above baselines for closeness and betweenness respectively. We outperform the baselines in almost all cases.} 
% We further strengthen our hypothesis that high degree nodes localized in inner most core of the network are also central nodes with respect to closeness and betweenness. We compare F-Score computed on two sets of predicted high central nodes, where the first set is composed of high degree inner most core nodes and the second set is generated by $k$\footnote{We have used k=10} random draws from the network  ignoring core numbers. Average F-Score of the latter prediction set is drastically poorer compared to its former counterpart as shown in Table~\ref{tab:r1}}

 \begin{table*}[!htb]
\centering
\caption{$F1$-score results for the predicted top-10 central nodes for closeness averaged over multiple temporal snapshots. Mean $(\mu)$ and std. dev. $(\sigma)$ of results are reported and the obtained results are compared against the existing baselines. The value of $r$ is set to $20$. The best results are marked in \textbf{bold}.}
\label{tab:r1}
\begin{adjustbox}{width=1\textwidth, height=1.35cm}\textcolor{black}{
\begin{tabular}{cccccc}
\hline 
{\bf Networks } & \begin{tabular}[c]{@{}c@{}}\{\bf $F1$-score Close $(\mu,\sigma)$\\ (Our method: High degree from core)\}\end{tabular} & \begin{tabular}[c]{@{}c@{}}\{\bf $F1$-score Close $(\mu,\sigma)$\\ (Global degree )\}\end{tabular} & \begin{tabular}[c]{@{}c@{}}\{\bf $F1$-score Close $(\mu,\sigma)$\\ (Uniform)\}\end{tabular} & \begin{tabular}[c]{@{}c@{}}\{\bf $F1$-score Close $(\mu,\sigma)$\\ (W1)\}\end{tabular} & \begin{tabular}[c]{@{}c@{}}\{\bf $F1$-score Close $(\mu,\sigma)$\\ (W2)\}\end{tabular} \\ 
 \hline
AS & \cellcolor{white}{\bf 0.81,0.08} & 0.24,0.08 &0.74,0.05 & 0.75,0.09 & 0.75,0.08 \\ 
CA & \cellcolor{white}{\bf 0.77,0.08} & 0.1,0.03 & 0.73,0.08 & 0.74,0.07 & 0.74,0.09   \\ 
HT  & \cellcolor{white}{\bf 0.42,0.3} & 0.12,0.08 & 0.19,0.03 & 0.18,0.12 & 0.14,0.09 \\ 
HP & \cellcolor{white}{\bf 0.46,0.29} & 0.07,0.08 & 0.23,0.14& 0.0,0.0 & 0.0,0.0 \\ 
SO & \cellcolor{white}{\bf 0.39,0.22} & 0.26,0.31 &0.21,0.15 & 0.21,0.15 & 0.23,0.15 \\ 
WT & \cellcolor{white}{\bf 0.31,0.19} & 0.15,0.11 & 0.25,0.16 & 0.26,0.15 & 0.26,0.15 \\  
FW  & \cellcolor{white}{\bf 0.24,0.19}  & 0.21,0.16 & 0.13,0.14& 0.07,0.11& 0.07,0.11 \\
SU & 0.02,0.21  & 0.007,0.19 & \cellcolor{white}{\bf 0.04,0.024} & 0.02,0.04 & 0.02,0.04  \\ 
\hline
\end{tabular}}
\end{adjustbox}
\end{table*}

 \begin{table*}[!htb]
\centering
\caption{$F1$-score results for the predicted top-10 central nodes for betweenness averaged over multiple temporal snapshots. Mean $(\mu)$ and std. dev. $(\sigma)$ of results are reported and the obtained results are compared against existing baselines. The value of $r$ is set to $20$. The best results are marked in \textbf{bold}.}
\label{tab:r2}
\begin{adjustbox}{width=1\textwidth, height=1.35cm}\textcolor{black}{
\begin{tabular}{cccccc}
\hline
{\bf Networks } & \begin{tabular}[c]{@{}c@{}}\{\bf $F1$-score Bets $(\mu,\sigma)$\\ (Our method:High degree from core)\}\end{tabular} & \begin{tabular}[c]{@{}c@{}}\{\bf $F1$-score Bets $(\mu,\sigma)$\\ (Global degree )\}\end{tabular} & \begin{tabular}[c]{@{}c@{}}\{\bf $F1$-score Bets $(\mu,\sigma)$\\ (Uniform)\}\end{tabular} & \begin{tabular}[c]{@{}c@{}}\{\bf $F1$-score Bets $(\mu,\sigma)$\\ (W1)\}\end{tabular} & \begin{tabular}[c]{@{}c@{}}\{\bf $F1$-score Bets $(\mu,\sigma)$\\ (W2)\}\end{tabular} \\
\hline
\vspace{0.7mm}
AS & \cellcolor{white}{\bf 0.72,0.08} & 0.24,0.08 &0.69,0.05 & 0.7,0.09 & 0.68,0.08 \\ 
CA & \cellcolor{white}{\bf 0.64,0.08} & 0.1,0.03 & 0.59,0.08 & 0.61,0.07 & 0.61,0.09   \\ 
HT  & \cellcolor{white}{\bf 0.46,0.3} & 0.12,0.08 & 0.08,0.03 & 0.14,0.12 & 0.14,0.09 \\ 
HP & \cellcolor{white}{\bf 0.52,0.29} & 0.07,0.08 & 0.16,0.18& 0.06,0.11 & 0.06,0.11 \\ 
SO & \cellcolor{white}{\bf 0.39,0.22} & 0.26,0.31 &0.32,0.19 & 0.21,0.15 & 0.28,0.15 \\ 
WT & \cellcolor{white}{\bf 0.31,0.19} & 0.15,0.11 & 0.29,0.16 & 0.26,0.15 & 0.3,0.15 \\  
FW  & \cellcolor{white}{\bf 0.2,0.19}  & 0.21,0.16 & 0.13,0.14& 0.07,0.11& 0.13,0.11 \\
SU & \cellcolor{white}{\bf 0.18,0.21}  & 0.07,0.19 & 0.14,0.024 & 0.02,0.04 & 0.11,0.17  \\ 
\hline
\end{tabular}}
\end{adjustbox}
\end{table*}

\subsection{Experimental evaluation of computational complexity}
\textcolor{black}{We also estimate the amount of time required to actually find the high central nodes (using traditional shortest-path based techniques) for the real world networks where the prediction accuracy ($F1$-score) is high. We report these results in Table~\ref{time}, which show that the time required to predict node ids with high centrality (inclusive of the time required to compute the classification parameters) is substantially low compared to actually finding them.} %\TODO{Soumya: Add machine specifications: CPU, RAM, OS etc.}
%\textcolor{red}{SB: Write whether the time included all the steps for both classification and prediction. If it only includes time for predict and not classfication, thenr eviewers might question how long does the clustering require.}
\begin{table}[!htb]
\centering
\caption {\label{time} Experimental evaluation of the computational complexity for prediction and parameter calculation. Evaluation was done on a workstation desktop running 64bit Ubuntu 14.04 with Intel Xeon E312xx family processor and 32GB RAM.}
\begin{adjustbox}{max width=0.7\textwidth}
\huge {\textcolor{black}{
\begin{tabular}{ c  c  c c  }
\hline
{\bf Network } & {\bf  Time (secs)} & {\bf Time (secs)} & {\bf Time (secs)} \\
& Traditional method & Classification parameters & Prediction method\\
\hline
AS  & 34.19 & 4.23  & 1 \\ 
CA  & 1946.2 & 5.67  & 2  \\ 
HP  & 271 & 4.89  & 1 \\ 
HT   & 84 & 3.11  & 1 \\ 
 \hline
 \end{tabular} }}
 \end{adjustbox} 
 \end{table}

\subsection{Results on synthetic networks}

\textcolor{black}{We further evaluate our classification and prediction frameworks on a set of 20 synthetic networks of different sizes, different core structures and different number of temporal snapshots generated using two  network generator tools -- Musketeer~\cite{gutfraind2015multiscale} and Dancer~\cite{benyahia2016dancer}. While $N5,N6,N7,N12$ have been generated by the Dancer tool, the remaining networks have been generated by the Musketeer tool. For all the 20 new networks, we report the number of temporal snapshots and the number of nodes and edges in the largest snapshot in Table~\ref{new:results}. For each network, we first obtain the cumulative distributions corresponding to all the four parameters $(EF,CFX,ED,CV)$ by estimating these values for a certain number of time steps. Since we already have the two well-defined clusters from the 8 real datasets and their corresponding centroids, we simply calculate the distance of the distribution using $D$-statistic from the two centroids and assign the network to the cluster corresponding to the nearest centroid (see Rocchio classification~\cite{baeza1999modern}). This scheme enables us to avoid re-clustering every time a new dataset is available.} 

\textcolor{black}{As a following step, likewise real world networks, here also we predict the overlap values and the exact nodes. We report the corresponding error percentages and $F1$-scores in Table~\ref{new:results}. Once again, as observed earlier, the results are best for the $GGGG$ class of networks.}
%We have also calculated the predicted overlap for both high centrality metrics between two temporal snaphots as well as $F1$-score for the actual high central nodes. We report the averaged predicted overlap and F-Score in Table~\ref{new:results}. Results obtain reinforces our hypothesis that for networks where maximum heuristics belong to category Good(G), the prediction scores are high. }

\begin{table}[!htb]
\centering
 \caption {\label{new:results} Results for the test suite of synthetic networks generated by the Dancer tool ($N5$, $N6$, $N7$, $N12$) and the Musketeer tool (remaining networks). Average error in predicted overlap as well as the mean $F1$-score of the predicted top-10 central nodes averaged over multiple temporal snapshots is reported in terms of mean$ (\mu)$ and std. dev. $(\sigma)$. Networks are grouped with respect to the class they belong to.} 
\begin{adjustbox}{max width=\textwidth,max height=3.5cm}
\huge {
\textcolor{black}{
\begin{tabular}{  c  c  c  c  c  c c  c  c  }
\hline 
\begin{tabular}[c]{@{ }c@{ }}
{\bf Network}\\
{ \bf Name}
\end{tabular} & {\bf Nodes} & {\bf Edges} & {\bf Time steps} &
\begin{tabular}[c]{@{ }c@{ }}
{\bf Bet. overlap $(\mu,\sigma)$}\\
{ \bf pred. err.}
\end{tabular}
   & \begin{tabular}[c]{@{ }c@{ }}
{\bf Close. overlap $(\mu,\sigma)$}\\
{ \bf pred. err.}
\end{tabular} & \begin{tabular}[c]{@{ }c@{ }}
{\bf $F1$-score $(\mu,\sigma)$}\\
{ \bf Bet.}
\end{tabular}& \begin{tabular}[c]{@{ }c@{ }}
{\bf F1 Score $(\mu,\sigma)$}\\
{ \bf Close.}
\end{tabular} & \begin{tabular}[c]{@{ }c@{ }}
{\bf Network}\\
{ \bf Category}
\end{tabular}  \\ 
\hline 
\vspace{0.8mm}
%\rowcolor{LightCyan}
$N1$ & 6437 & 16936 & 720 & 14.71, 13.38 & 17.78,17.31 & 0.73,0.10 & 0.82,0.09 & $GGGG$ \\ 
%\rowcolor{LightCyan}
$N2$ & 4432& 8462 & 720 & 13.78,12.22 & 14.67,13.75  & 0.79,0.09 & 0.81,0.08 & $GGGG$ \\ 
%\rowcolor{LightCyan}
$N3$ & 39545 & 12290 & 120 & 24.40,18.07 & 28.39,20.95 & 0.78,0.09 & 0.67,0.11 & $GGGG$\\ 
%\rowcolor{LightCyan}
$N4$ & 22871 & 44566 & 120 & 26.22,17.51 & 27.19,19.09 & 0.92,0.07 & 0.75,0.10 & $GGGG$\\
%\rowcolor{LightCyan}
$N5$ & 2782 & 170037 & 50 & 10.28,6.67 & 12.6,8.73 & 0.89,0.08 & 0.94,0.01 &  $GGGG$\\
%\rowcolor{Gray}
$N6$ & 1315 & 5136 & 50 & 10.54,4.62 & 13.17,6.19 &  0.54,0.09 & 0.71,0.13 &  $GGGB$\\
%\rowcolor{Gray}
$N7$ & 10316 & 67949 & 50 & 34.11,25.52 & 15.57,19.97 & 0.44,0.26 & 0.82,0.26 &  $GGGB$\\
%\rowcolor{Gray}
$N8$ & 25801 & 301995 & 120 & 22.09,24.15 & 22.26,21.35 & 0.21,0.21& 0.31,0.25 & $GGBG$ \\ 
%\rowcolor{Gray}
$N9$ & 21021 & 204460  & 120 & 18.74,19.08 & 28.54, 25.72 & 0.22 ,0.20& 0.34,0.29 & $GGBG$ \\  
%\rowcolor{Gray}
$N10$ & 15662 & 128463 & 82 & 30.45,24.71 & 41.68,30.40 & 0.38,0.26 & 0.45,0.29  & $GGBG$ \\ 
%\rowcolor{Gray}
$N11$ & 12832 & 91984 & 82 & 33.88,24.81 & 34.97,26.03 & 0.4,0.28& 0.47,0.33  & $GGBG$\\
%\rowcolor{orange}
$N12$ & 3525 & 194655 & 50 & 21.65,14.25 & 27.61,10.91 & 0.58,0.09 & 0.37,0.12 &  $GGBB$\\
%\rowcolor{ofwhite}
$N13$ &  207177 & 624796 & 90 & 30.09,27.90 & 43.63,30.33 & 0.56,0.11 & 0.45,0.12 & $BBBG$ \\ 
%\rowcolor{ofwhite}
$N14$ & 164574 & 438655 & 90 & 35.09, 27.90 & 43.6,30.33& 0.41,0.18 &  0.39,0.12 &  $BBBG$\\ 
$N15$ & 73499 & 166774 & 65 & 21.82,28.35 & 24.33,26.28 & 0.28,0.15 & 0.31,0.02  & $BBBB$\\ 
$N16$ & 55681 & 111188 & 65 & 35.18,30.89 & 11.90, 29.83 & 0.33,0.18 & 0.36,0.19 & $BBBB$\\  
$N17$ & 19591 & 28431 & 48 & 112.34,75.96 & 119.90,138.66 & 0.22,0.24 & 0.28,0.27 & $BBBB$ \\ 
$N18$ & 15403 &  20063 & 48 & 99.61,70.48 & 78.68,62.60 & 0.19,0.23& 0.23,0.27 & $BBBB$\\ 
$N19$ & 8472 & 14107  & 90 & 52.54,70.68 & 78.58, 62.60 & 0.27,0.16& 0.13,0.19 & $BBBB$\\ 
$N20$ & 6758 & 9895 & 90 & 51.54,70.69 & 50.59,76.21 & 0.31,0.19 & 0.12,0.21 &  $BBBB$\\
\hline
 \end{tabular} }}
 \end{adjustbox} 
 \end{table}

\section{Validation}\label{valid}

In this section we demonstrate that the behavior of our predicted high centrality vertices is very similar to the actual high centrality vertices in a practical context. 
We select one representative example from two extreme classes of networks (AS, category $GGGG$ and WT, category $BBBB$) and show that for those which conform well to our hypothesis, the predicted high central vertices and the actual high central vertices behave similarly, while for those which do not conform well any random selection of vertices behaves similarly to the actual high centrality vertices.

\subsection{Validation for closeness centrality}

\begin{figure}[!htb]
    \centering
    \begin{minipage}{.5\textwidth}
        \centering
        \includegraphics[width=\linewidth, height=0.17\textheight]{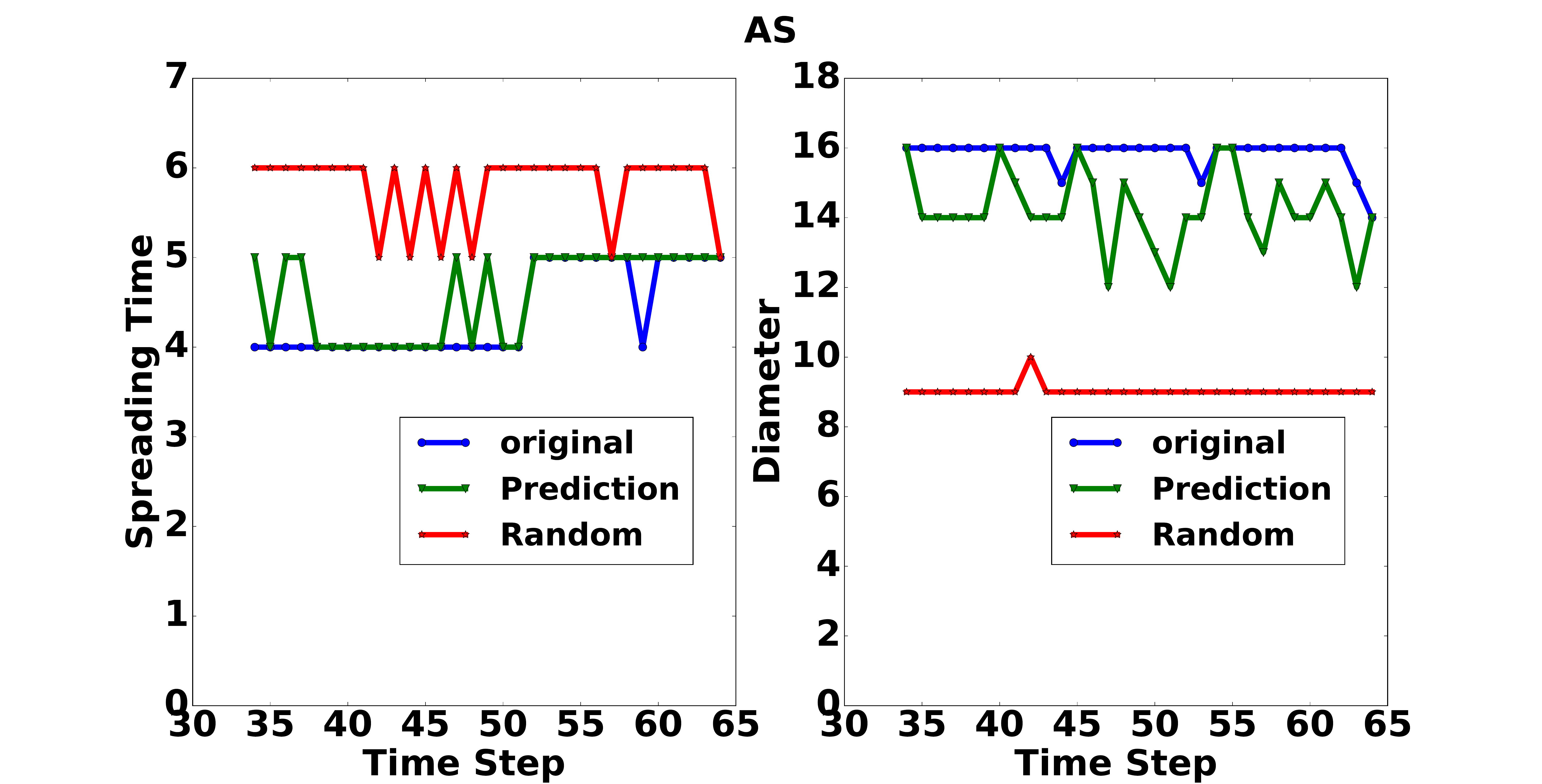}
        \caption{AS network}
        \label{fig:As2}
    \end{minipage}%
    \begin{minipage}{0.5\textwidth}
        \centering
        \includegraphics[width=\linewidth, height=0.17\textheight]{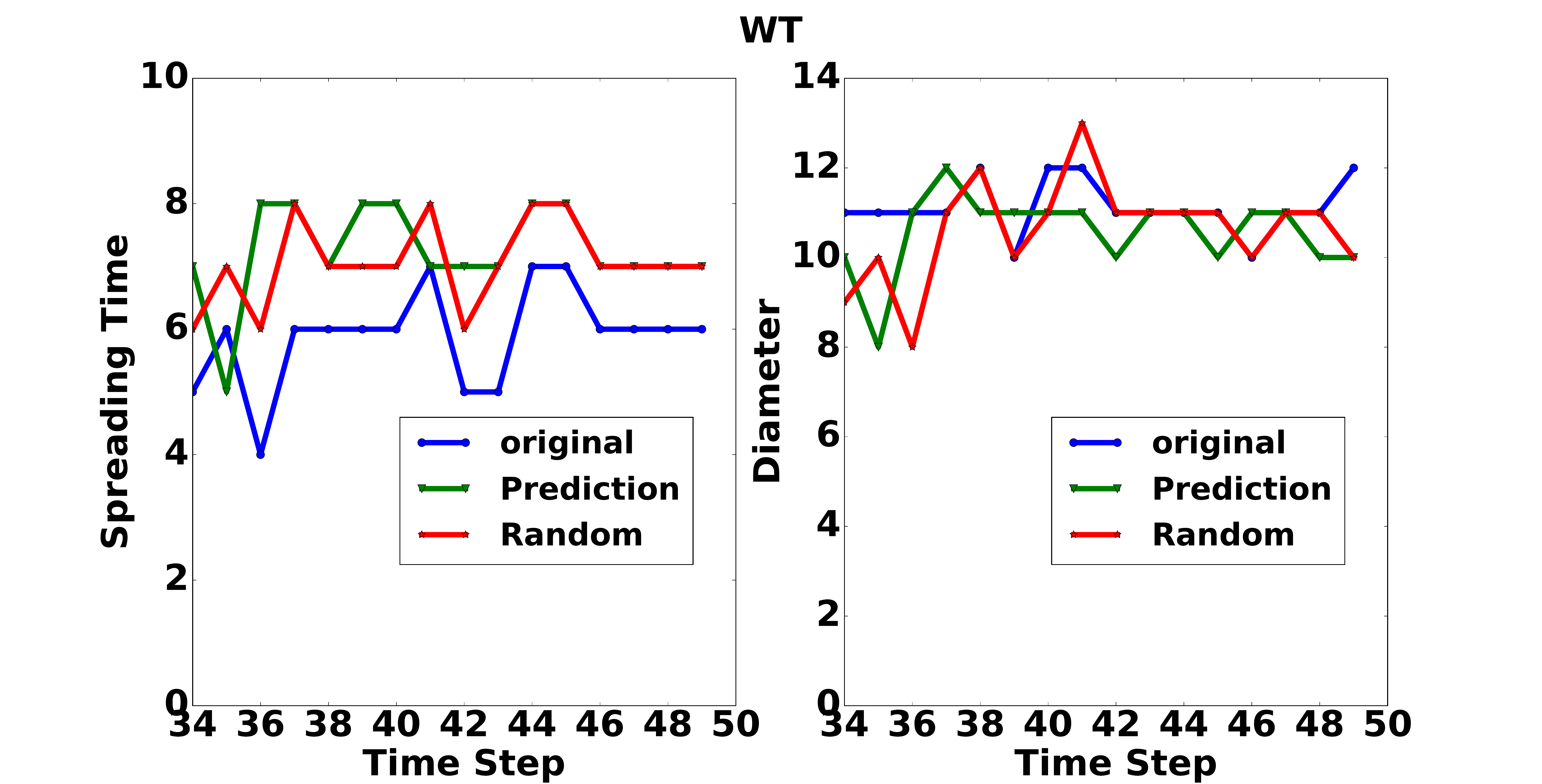}
        \caption{WT network}
        \label{fig:As2}
    \end{minipage}
     \label{exp}  \caption{\label{fig:Spread} The left panel shows  validation results for AS network and the right panel for the WT network. Left: Time for spreading a message with high closeness centrality vertices as initial seeds. Right (betweenness): The diameter size after removing high betweenness centrality vertices. Color online.}
\end{figure}

To validate for the closeness centrality, we obtain the top 10 high central vertices, the top 10 predicted high central vertices and 10 random vertices from the graph $G_{t}$, the $t^\textrm{th}$ snapshot in the  time series. We use the vertices from each set as seeds to disseminate a message in the network. At each iteration a seed vertex $u$ propagates its message to all its neighbors, and these new vertices who just received the message is appended into the seed set. We stop the iteration when all the vertices have received the message. 
We perform this experiment for all the networks in the time series and report the results in Figure~\ref{fig:Spread}.  

For the {\em AS network}, the actual top central vertices are the fastest propagator of the message in the network, and this trend remains consistent in all the time steps. Our predicted top central vertices almost always show similar behavior to the actual vertices. The  randomly selected vertices, in contrast, take longer to spread the message. For the {\em WT network}, however the trend from the actual, the predicted, and the random vertices are very similar qualitatively (and quantitatively as well) to one another.

\subsection{Validation for betweenness centrality} 

In order to validate for betweenness centrality we posit that since high fraction of shortest paths should pass through the high betweenness vertices, removing them would increase the diameter by a significant margin. We obtain the actual top 10 betweenness centrality vertices, the top 10 predicted vertices and a set of 10 random vertices. In each case, we remove these vertices from the network and calculate the diameter.

For the {\em As network}, the removal of the actual top 10 betweenness centrality vertices leads to the larger diameter and the size is very similar to that obtained for the predicted high betweenness vertices. Removing random vertices affects the diameter the least and is much lower than the other two above cases. For the {\em WT network},  the effects on the diameter due to removal of the actual top, the predicted top as well as the random vertices are the same. This  indicates that the betweenness centrality values in a network like WT are very uniform across vertices and therefore selecting high betweenness central vertices are of not much advantage in the practical context.

\section{Theoretical rationale using core connectedness} \label{proof_sketch}

We introduce the \textit{core connectedness} (CC) property to formalize our hypothesis and to quantitatively distinguish between the networks whose high centrality vertices are in a small and dense innermost core from those where this property does not hold.

\noindent{\bf Defining core connectedness}: Consider a graph $G$ with a core-periphery structure. The shells are consecutively numbered as $S_{1}, \ldots, S_i, S_{i+1}, \ldots, S_{max}$, where $S_1$ is the outermost shell and $S_{max}$ is the innermost shell. For ease of expression we will use shell numbers interchangeably with their integer values, i.e. $S_i-S_j$ to denote the difference $i-j$.

We define a path between two specified vertices as a sequence of alternating vertices and edges where no vertex is repeated.  The shortest path (distance) is the smallest such sequence. We denote the length of the distance between two vertices $v$ and $u$ as $P_{v \rightarrow u}$. If there is no path between $v$ and $u$ then $P_{v \rightarrow u}=\infty$.
For each pair of vertices $v$ and $u$ we obtain the following two paths; {\bf(i)} The shortest path between $v$ and $u$, with the constraint that the sequence contains at least one vertex from $S_{max}$. The length of this path is denoted as $P_{v \rightarrow u}^{max}$. {\bf(ii)} The shortest path between $v$ and $u$, with the constraint that the sequence does not contain any  vertex from $S_{max}$. The length of this path is denoted as $P_{v \rightarrow u}^{O}$. Let the length of the shortest path between $v$ and $u$, without any constraints be $P_{v \rightarrow u}^{X}$.

Given these definitions at least one of the paths lengths , $P_{v \rightarrow u}^{O}$ or $P_{v \rightarrow u}^{max}$ (but perhaps not both) would be equal to $P_{v \rightarrow u}^{X}$.  

We  define {\em a core connected (CC) network is one where $P_{v \rightarrow u}^{max} \le P_{v \rightarrow u}^{O} $ for $v,u \in V$,  $(v,u) \not\in E$ and $P_{v \rightarrow u}^{max} \neq \infty$.}

\begin{figure}[htb]
\centering
\includegraphics[scale=.3]{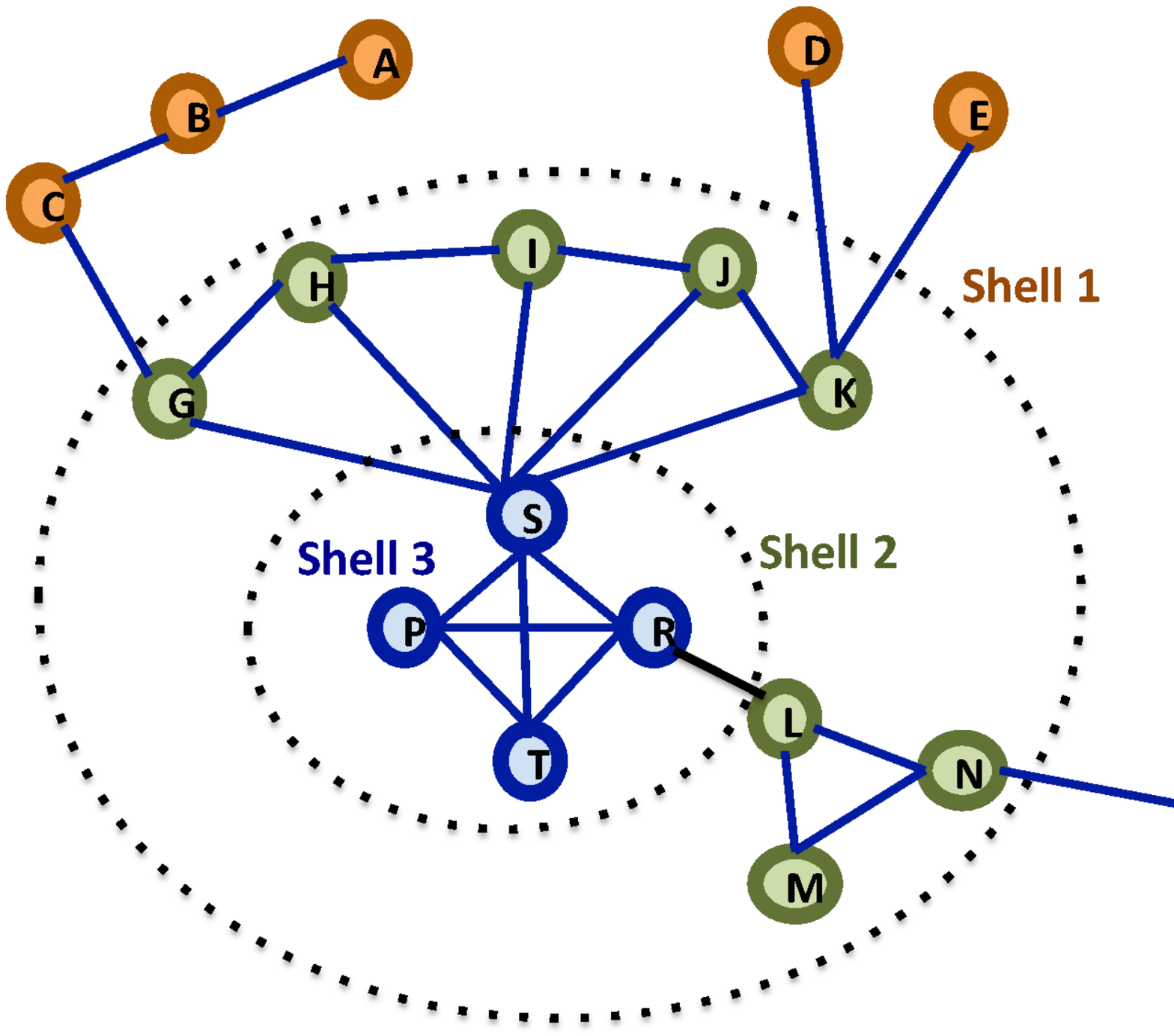} 
\caption{\label{fig:CCC} {\bf Example of a core connected network.}  (color online) The network has three shells. Lengths of paths between all non-neighboring vertices that pass through shell 3 are also the absolute shortest between them. Example $P_{C \rightarrow E}^{max}$=4 while $P_{C \rightarrow E}^{O}$=6. Color Online.}
\end{figure}

In other words, in a core connected network, if two non neighboring vertices $v$ and $u$, have a path through the innermost core then that path is the shortest path between them. An example of a core connected network is given in Figure~\ref{fig:CCC}.

For networks that are core connected, the CC strength of a network can be computed as the number of paths that pass through the core to the total number of paths. It is easy to see that the higher the CC strength, the higher the likelihood that path-based high centrality vertices will be in the innermost core.

\vspace{-1mm}
\section{Conclusion and future work}

In this paper, we present a two-step algorithm for predicting the high centrality vertices in time varying networks that alleviates the need for the costly shortest path computations in each temporal snapshot. In the first step, we predict the overlap between the set of high closeness and betweenness centrality vertices of the previous time steps to the set of high centrality vertices of the future time steps. In the second step, once the current network snapshot is available, we further analyze its innermost core to find the exact ids of the high centrality vertices. To the best of our knowledge, {\em this is the first algorithm to predict the exact ids of high centrality vertices}.

In order to perform accurate predictions, we hypothesize that in real world time-varying networks, a majority of high central nodes reside in the innermost core. Based on this hypothesis, we classify an array of real-world networks into those that conform strongly to this hypothesis and those that do not. Our predictions of the high centrality nodes are remarkably accurate for the networks that strongly conform to our hypothesis. We further validate our predictions through two different appropriate applications for the two different centrality measures. Finally, we analytically demonstrate the CC property which is the key idea behind why our method works.
\if{0}
Some of our important contributions are as follows: {\em (i)} introducing the CC property and the analytical justification proving that the core periphery structure of a network can be used to identify top central nodes, {\em (ii)} an algorithm based on time series forecasting to predict the overlap of the top central nodes across time points {\em (iii)} an  algorithm to exactly pinpoint the ids of the nodes that have high centrality when the network is available in time and {\em (iv)} a detailed validation scheme to show that the behavior of the high centrality nodes predicted  by our algorithm indeed matches the behavior of the actual high centrality nodes.\fi

In future, we would like to extend the prediction scheme for other types of centrality measures, beyond shortest-distance based entities (e.g., eigenvector and spectral centrality). Further, we would like to propose efficient generative models of temporal networks that preserve the CC property. We also plan to investigate if networks that strongly conform to the CC property are more resistant to noise/failures. Finally, we would like to explore if the CC property can be used to solve other known difficult problems in network science like community detection. Most of the codes used for experiments are made available at \url{https://github.com/Sam131112/temporal} to promote reproducible research.

\bibliographystyle{spmpsci}    
\bibliography{sigproc}   
\end{document}